\documentstyle[12pt,aaspp4]{article}

\newcommand\about     {\hbox{$\sim$}}

\def\E#1{\hbox{$10^{#1}$}}

\def\case#1/#2{\hbox{$\frac{#1}{#2}$}}
\def\about  {\hbox{$\sim$}}

\def\O                {\hbox{$O$}}
\def\E                {\hbox{$E$}}
\def\Oa               {\hbox{$O_a$}}
\def\Ea               {\hbox{$E_a$}}
\def\Jg               {\hbox{$J_g$}}
\def\Fg               {\hbox{$F_g$}}
\def\J                {\hbox{$J$}}
\def\F                {\hbox{$F$}}
\def\N                {\hbox{$N$}}

\def\comm#1           {\tt #1}
\def\refto#1          {\ref #1}

\begin{document}


\title{    Variable Faint Optical Sources Discovered by  
                Comparing POSS and SDSS Catalogs            }

\author{
Branimir Sesar\altaffilmark{\ref{Zagreb},\ref{Princeton}}, 
Domjan Svilkovi\'{c}\altaffilmark{\ref{Zagreb}},
\v{Z}eljko Ivezi\'{c}\altaffilmark{\ref{Princeton},\ref{RF}} 
Robert H. Lupton\altaffilmark{\ref{Princeton}},
Jeffrey A. Munn\altaffilmark{\ref{USNO2}},
Douglas Finkbeiner\altaffilmark{\ref{Princeton},\ref{HF}},
William Steinhardt\altaffilmark{\ref{Princeton}},
Rob Siverd\altaffilmark{\ref{Princeton}},
David E. Johnston\altaffilmark{\ref{Princeton}},
Gillian R. Knapp\altaffilmark{\ref{Princeton}},
James E. Gunn\altaffilmark{\ref{Princeton}},
Constance M. Rockosi\altaffilmark{\ref{UW}},
David Schlegel\altaffilmark{\ref{Princeton}},
Daniel E. Vanden Berk\altaffilmark{\ref{Pittsburgh}},
Pat Hall\altaffilmark{\ref{Princeton},\ref{Catolica}},
Donald P. Schneider\altaffilmark{\ref{PennState}},
Robert J. Brunner\altaffilmark{\ref{Illinois}}
}

\newcounter{address}
\setcounter{address}{1} 
\altaffiltext{\theaddress}{University of Zagreb, Dept. of Physics, Bijeni\v{c}ka 
cesta 32, 10000 Zagreb, Croatia \label{Zagreb}} 
\addtocounter{address}{1}\altaffiltext{\theaddress}{Princeton University 
Observatory, Princeton, NJ 08544 \label{Princeton}}
\addtocounter{address}{1}\altaffiltext{\theaddress}{H.N. Russell Fellow, on leave from 
             the University of Washington \label{RF}}
\addtocounter{address}{1}\altaffiltext{\theaddress}{U.S. Naval Observatory,
Flagstaff Station, P.O. Box 1149, Flagstaff, AZ 86002 \label{USNO2}}
\addtocounter{address}{1}\altaffiltext{\theaddress}{Hubble Fellow \label{HF}}
\addtocounter{address}{1} \altaffiltext{\theaddress}{University of Washington,
Dept. of Astronomy, Box 351580, Seattle, WA 98195 \label{UW}}
\addtocounter{address}{1}
\altaffiltext{\theaddress}{University of Pittsburgh, Dept. of Physics \& Astronomy,
3941 O'Hara Street, Pittsburgh, PA 15260
\label{Pittsburgh}}
\addtocounter{address}{1} 
\altaffiltext{\theaddress}{Pontificia Universidad Cat\'{o}lica de Chile, Departamento de 
Astronom\'{i}a y Astrof\'{i}sica, Facultad de F\'{i}sica, Casilla 306, Santiago 22, Chile
\label{Catolica}}
\addtocounter{address}{1} \altaffiltext{\theaddress}{Department of Astronomy
and Astrophysics, The Pennsylvania State University, University Park, PA 16802
\label{PennState}}
\addtocounter{address}{1} \altaffiltext{\theaddress}{NCSA and Department of 
      Astronomy, University of Illinois, Urbana, IL 61801  \label{Illinois}}

\begin{abstract}
We present a study of variable faint optical sources discovered 
by comparing the Sloan Digital Sky Survey (SDSS) and the Palomar
Observatory Sky Survey (POSS) catalogs. We use SDSS measurements 
to photometrically recalibrate several publicly available POSS 
catalogs (USNO-A2.0, USNO-B1.0, GSC2.2 and DPOSS). A piecewise 
recalibration of the POSS data in 100 arcmin$^{2}$ patches (one 
SDSS field) generally results in an improvement of photometric 
accuracy (rms) by nearly a factor of two, compared to the original 
data. In addition to the smaller core width of the error distribution, 
the tails of the distribution become much steeper after the 
recalibration. These improvements are mostly due to the very dense 
grid of calibration stars provided by SDSS, which rectifies the 
intrinsic inhomogeneities of Schmidt plates. We find that the 
POSS I magnitudes can be improved to $\sim$0.15 mag accuracy, and 
POSS II magnitudes to $\sim$0.10 mag accuracy. The smallest final 
errors are obtained with the GSC2.2 catalog for which they 
approach 0.07 mag at the bright end.

We use the recalibrated catalogs for the $\sim$2,000 deg$^2$ 
of sky in the SDSS Data Release 1 to construct a catalog of 
$\sim$60,000 sources variable on time scales 10-50 years. We find 
that at least 1\% of faint optical sources appear variable at the 
$>$0.2 mag level, and that about 10\% of the variable population 
are quasars, although they represent only 0.25\% of all point 
sources in the adopted flux-limited sample. A series of statistical 
tests based on the morphology of SDSS color-magnitude and color-color 
diagrams, as well as visual comparison of images and comparison with 
repeated SDSS observations, demonstrate the robustness of the selection 
methods. Candidate variable sources are correctly identified in more 
than 95\% of cases, and the fraction of true variables among the 
selected candidates is as high as 73\%. We quantify the distribution 
of variable sources in the SDSS color-color diagrams, and the 
variability characteristics of quasars. The observed long-term variability 
(structure function) is smaller than predicted by the extrapolation 
of the power-law measured for short time scales using repeated SDSS 
imaging (0.35 vs. 0.60 mag for SDSS-POSS I, and 0.24 vs. 0.35 mag for 
SDSS-POSS II, rms). This turn-over in structure function suggests 
that the characteristic time scale for quasar variability is of the 
order one year. The long-term ($\ga$1 year) quasar variability 
decreases with luminosity and rest-frame wavelength similarly to the 
short-term ($\la$1 year) behavior. We also demonstrate that candidate 
RR Lyrae stars trace the same halo structures, such as the Sgr dwarf 
tidal stream, that were discovered using repeated SDSS observations. 
We utilize the POSS-SDSS selected candidates to constrain the halo 
structure in the parts of sky for which repeated SDSS observations do 
not exist.
\end{abstract}


\section{Introduction }

The time domain represents another dimension, in addition to spectral and spatial, 
in the exploration of celestial objects. Despite the importance of variability 
phenomena, the properties of optically faint variable sources are by and large
unknown. There are about 10$^9$ stars brighter than $V=20$ in the sky,
and at least 3\% of them are expected to be variable at a few percent
level (Eyer, 1999). However, the overwhelming majority are not recognized
as variables even at the brightest magnitudes: 90\% of variable
stars with $V<12$ remain to be discovered (Paczy\'{n}ski, 2000).
Paczy\'{n}ski (1996) lists striking examples of the serious incompleteness 
in the available samples of variable stars: eclipsing binaries of Algol type, 
and contact binaries (W UMa stars) are incomplete fainter than $V \about 12$,
and RS CVn type binaries are complete only to $V \about 5$. Another vivid example 
of serious selection effects is the sky distribution of RR Lyrae stars: objects 
listed in the 4$^{th}$ General Catalogue of Variable Stars (the main resource for
variable stars) are distributed in isolated square patches with the 
size and shape of the Schmidt plates used to discover them.

The discrepancy between the utility of variable stars and the available
observational data has prompted several contemporary projects aimed at regular monitoring 
of the optical sky. The current state-of-the-art also greatly benefited from past 
and present microlensing searches (Paczy\'{n}ski, 2001). Some of the more prominent 
surveys in terms of the sky coverage, depth, and cadence are:

\begin{itemize}
\item
The Faint Sky Variability Survey (Huber, Howell, Everett \& Groot, 2002)
is a very deep ($V < 24$) $BVI$ survey of 23 deg$^2$ of sky, containing
about 80,000 sources
\item
The QUEST Survey (Vivas et al. 2001) monitors 600 deg$^2$ of sky
to a limit of $V=21$.
\item
ROTSE (Akerlof et al. 2003) monitors the entire observable sky twice
a night to  a limit of $V=15.5$.
\item
OGLE (most recently OGLE III; Udalski et al. 2002) monitors \about 100 deg$^2$ 
towards the Galactic bulge to a limit of $I=20$. Due to a very high stellar 
density towards the bulge, OGLE II has detected over 200,000 variable stars 
(Wo\'{z}niak et al. 2002).
\end{itemize}

These and other surveys have demonstrated that in addition to variable stars,
there are many other exciting photometrically variable objects in the sky. For 
example, ROTSE~I detected an optical flash generated by a gamma ray burst at 
a redshift of 1.6, the most luminous optical source ever measured ($V=9$, 
$M_V=-36.4$, Vestrand et al. 2002). The detection of such optical flashes 
may place strong constraints on the physical mechanisms responsible for gamma 
ray bursts. Similarly, the variability of quasars offers significant clues for 
the origin of their emission (e.g. Trevese, Kron \& Bunone, 2001).

Recognizing the outstanding importance of variable objects, the
last Decadal Survey Report highly recommended a major new initiative for
studying the variable sky, the Large Synoptic Survey Telescope (LSST).
The LSST\footnote{There are currently two designs considered for implementation:
a distributed aperture approach (Pan-STARRS, Kaiser et al. 2002) and a single 
large-aperture telescope (Tyson 2002).} will offer an unprecedented view of the faint 
variable sky: according to the current designs it will scan the entire accessible sky 
every three nights to a limit of $V\about 24$. Compared to any other survey currently
available, the data from LSST will be revolutionary. Yet, at least
half a decade or more will elapse before the first photons are detected by the LSST.
Meanwhile, the already available Palomar Digital Sky Surveys (POSS I and POSS II, for
references see Appendix A) and ongoing Sloan Digital Sky Survey (see Section 2.2) 
can be used to study the properties of faint ($r \sim 20$) optical sources, and here 
we present such a study.

The comparison of the POSS and SDSS surveys allows studies of the long term variability 
with time scales of up to half a century. By necessity, such studies are based on a 
small number of observations of the same objects to constrain the 
ensemble properties of a sample of sources, as opposed to studying well-sampled light 
curves for a small number of objects. The lack of detailed information for individual 
objects is compensated in some ways by the large sample size. In addition, the 5-band 
accurate SDSS photometry can be used for very detailed source classification; e.g. separation 
of quasars and stars (Richards et al. 2002), spectral classification of stars to within 
1-2 spectral subtypes (Lenz et al. 1998, Finlator et al. 2000, Hawley et al. 2002), 
and even remarkably efficient color selection (thanks to accurate $u$ band photometry) of the
low-metallicity G and K giants (Helmi et al. 2002) and horizontal branch stars 
(Yanny et al. 2000, Ivezi\'{c} 2003a).

However, when using only a few epochs of observations, the robustness of variability detection critically 
depends on the stability of the photometric errors. While the SDSS photometric errors are 
small ($\sim$0.02) and well behaved (Ivezi\'{c} et al. 2003b), older photographic POSS data 
can have large errors (tenths of mag) due to intrinsic inhomogeneities of Schmidt plates and 
the lack of a sufficient number of calibration stars. This problem can be alleviated
to some extent by using photometric measurements of stars in the SDSS to recalibrate POSS 
catalogs. In Section 2 we describe such a recalibration method and demonstrate that
photometric errors in the POSS catalogs can be decreased by a factor of $\sim$2 (rms), 
with a significant improvement in the behavior of the error distribution tails. 
In Section 3, we use SDSS data and recalibrated POSS catalogs to select variable objects 
in $\sim$2000 deg$^2$ of sky from the SDSS Data Release 1 (Abazajian et al. 2003). 
In the same section we discuss tests which demonstrate the robustness of the selection 
algorithm, and quantify the distribution of variable sources in the SDSS color-color 
diagrams. The Milky Way halo structure traced by selected candidate RR Lyrae stars is 
discussed in Section 4, and in Section 5 we analyze the variability of quasars. 
Our main results are summarized in Section 6.

\section{ The Photometric Recalibration of POSS Catalogs using SDSS Measurements  }

\subsection{              The Input POSS Catalogs                          }

We utilize several publicly available POSS catalogs: USNO-A2.0, USNO-B1.0, GSC2.2
and DPOSS. A description of each catalog and references are listed in Appendix A. 
Here we briefly mention that all four catalogs utilize the same POSS I and POSS II 
Schmidt plates. However, the scanning and calibration procedures are different, and the 
source parameters, such as magnitudes, reported in different catalogs in general are 
not the same for the same sources detected on the same plates. USNO-A2.0 reports \O\ 
and \E\ magnitudes, hereafter \Oa\ and \Ea\ to distinguish them from \O\ and \E\ magnitudes 
reported in the USNO-B1.0 catalog. The latter catalog also lists $J$, $F$ and $N$ magnitudes.
The GSC2.2 catalog lists $J$ and $F$ magnitudes,  hereafter \Jg\ and \Fg\ to distinguish 
them from \J\ and \F\ magnitudes reported in the USNO-B1.0 catalog. The DPOSS 
catalog is also based on photographic $J,F$ and $N$, but they are
calibrated and reported as $G,R$ and $I$ magnitudes.
   
The completeness of the USNO-B1.0 catalog, measured using SDSS data, is discussed by 
Munn et al. (2004). Our analysis of other catalogs confirms their result that, in general,
POSS catalogs are $\sim$95\% complete at magnitudes brighter than 19--20 (depending
on a particular band/catalog), and have faint limits (which we define as the magnitude
where fewer than 50\% of SDSS sources are found in a POSS catalog) at $m\sim20.5-21$.

\subsection{              Sloan Digital Sky Survey           }

The SDSS is a digital photometric and spectroscopic survey which will cover up to one quarter 
of the Celestial Sphere in the North Galactic cap, and produce a smaller area ($\sim$ 
225 deg$^{2}$) but much deeper survey in the Southern Galactic hemisphere\footnote{See also 
http://www.astro.princeton.edu/PBOOK/welcome.htm} (\cite{York}, Stoughton et al. 2002,
Abazajian et al. 2003). The flux densities of detected objects are measured almost simultaneously 
in five bands ($u$, $g$, $r$, $i$, and $z$) with effective wavelengths of 3540 \AA, 
4760 \AA, 6280 \AA, 7690 \AA, and 9250 \AA\ (Fukugita et al. 1996, Gunn et al. 1998, 
Smith et al. 2002, Hogg et al. 2002).
The completeness of SDSS catalogs for point sources is $\sim$99.3\% at the bright end (Ivezi\'{c} 
et al. 2001), and drops to 95\% at limiting magnitudes\footnote{These values are determined by 
comparing multiple scans of the same area obtained during the commissioning year. Typical seeing 
in these observations was 1.5$\pm$0.1 arcsec.} of 22.1, 22.4, 22.1, 21.2, and 20.3 (the SDSS
saturation limit is $\sim$14 in the $r$ band, and somewhat brighter in other bands). 
All magnitudes are given on the 
AB$_{\nu}$ system (Oke \& Gunn 1983, for additional discussion regarding the SDSS photometric system 
see \cite{F96} and Fan 1999). The survey sky coverage of about $\pi$ steradians (10,000 deg$^{2}$)
will result in photometric measurements to the above detection limits for about 100 million stars
and a similar number of galaxies. Astrometric positions are accurate to about 0.1 arcsec per 
coordinate for sources brighter than $r\sim$20.5$^{m}$ (Pier et al. 2003), and the morphological 
information from the images allows robust star-galaxy separation to $r\sim$ 21.5$^{m}$ (Lupton 
et al. 2003). More technical details may be found in Stoughton et al. (2002), and on the
SDSS web site (http://www.sdss.org).

In this work we use the SDSS Data Release 1, which provides data for 2099 deg$^2$ of the sky.
The equatorial Aitoff projection of this area can be found at the SDSS web site (also Fig. 5 
in Ivezi\'{c} et al. 2003a).

\subsection{    Photometric Transformations Between POSS and SDSS Systems }
\label{phototrans}

We chose to synthesize magnitudes in the POSS bands using SDSS measurements, and
then recalibrate POSS catalogs using their original bands. The alternative of 
recalibrating POSS catalogs directly to the SDSS system is less desirable because
colors at the POSS epoch are poorly known, and this may have an effect on the 
photometric accuracy for variable sources. Following Monet et al. (2003), we 
adopt the following form to define synthetic POSS magnitudes, $m_{SDSS}$, calculated
from SDSS photometry:
\begin{equation}
\label{transEq}
              m_{SDSS}=m+b\,\,{\textrm{color}}+c
\end{equation}
where $m=g,r,g,r,i$ and color = $g-r$, $g-r$, $g-r$, $g-r$, $r-i$ for $O,E,J,F$, and $N$, 
respectively (e.g. $O_{SDSS}=g+b(g-r)+c$). Utilizing data for about $\sim$300 deg$^2$ of sky (SDSS runs 752 and 756),
we derived the best-fit values of coefficients $b$ and $c$ for each band and the POSS catalog.
We used only ``good'' sources defined as:

\begin{enumerate} 
\item Sources must be unresolved in SDSS data (note that the SDSS star-galaxy 
  separation is robust to at least $r^{*}\sim21.5$, which is significantly fainter 
  than the faint limit of the resulting sample). 
\item Sources must be isolated in SDSS data. This condition ensures that the 
  USNO/GSC/DPOSS photometry is not affected by difficult to measure, blended objects
\item Sources must not be saturated in $g$ or $r$ band in the SDSS data (roughly
  equivalent to $g,r > 14^{m}$), and must have $g < 19$.
\item The USNO/GSC/DPOSS and SDSS positions must agree to better than $2\arcsec$.
  This limit corresponds to a $\sim5\sigma$ cut on astrometric errors (Pier et al. 2003). 
\item The sources must have $u-g>0.7$ (measured by SDSS) to avoid highly variable
  quasars (see Section \ref{varqso}). 
\end{enumerate}

The best-fit values of coefficients $b$ and $c$, and the residual rms scatter
(which is a good measure of the mean photometric accuracy of the POSS catalogs) 
are listed in Table 1. Note that \J\ and \F\ magnitudes from the GSC2.2 catalog have the 
smallest residual scatter, while the \O\ and \E\ magnitudes have the largest scatter.

Similar values for $b$ and $c$ coefficients were derived for the USNO-B1.0 magnitudes 
by Monet et al. (2003). We have verified that adopting their transformations results in 
only slightly larger residual scatter for the other three catalogs. Thus, in order to prevent 
proliferation of various SDSS-POSS transformations, we adopt their transformations, which 
we list below for completeness, in the rest of this work (for DPOSS G and R 
magnitudes\footnote{We use upper case letters for DPOSS magnitudes to distinguish them
from SDSS $g$ and $r$ magnitudes.}, we use coefficients listed in Table 1).

\begin{eqnarray}
\label{synthetic}
\nonumber
            O_{\rm SDSS} = g+0.452\,(g-r)+0.08 \\
\nonumber
            E_{\rm SDSS} = r-0.086\,(g-r)-0.20 \\
\nonumber
            J_{\rm SDSS} = g+0.079\,(g-r)+0.06 \\
\nonumber
            F_{\rm SDSS} = r-0.109\,(g-r)-0.09 \\
\nonumber
            N_{\rm SDSS} = i-0.164\,(r-i)-0.44 \\
\end{eqnarray}

\subsection{                  The Recalibration Method                    }

The first basic premise of the recalibration method employed here is that the SDSS photometric
errors are negligible compared to errors in the POSS catalogs -- the SDSS photometric errors
are $\sim$0.02 mag, as demonstrated by repeated scans (Ivezi\'{c} et al. 2003b), while
the errors in the POSS catalogs are 0.1 mag or larger. The second premise is that
not more than a few percent of faint stars vary by more than a few percent, in agreement
with the available data and models (Eyer 1999). The third assumption is that
systematic errors are a significant contribution to photometric errors in the POSS
catalog, and thus can be calibrated out using a dense grid of calibration stars provided
by SDSS. We demonstrate empirically that indeed the accuracy of POSS-based photometry can 
be improved by about a factor of 2 in all analyzed catalogs.

The recalibration of the POSS catalogs is performed in two steps. In the first step the
subsamples of ``good'' objects (see Section \ref{phototrans}) are grouped by Schmidt plate 
and SDSS fields. One SDSS field has an area of 0.034 deg$^2$; this is sufficiently large to include
enough calibration stars (typically 50-200), and yet sufficiently small that the response of 
the Schmidt plates is nearly constant, as shown by Lattanzi \& Bucciarelli (1991). To avoid edge
effects, we use a running window with the width of 3 SDSS fields (0.45$^\circ$, see the next
section for more details). 

For each of the five POSS magnitudes, we minimize $\sum(m_{recalib}-m_{SDSS})^{2}$
using the least-square method, where 
\begin{equation}
                m_{recalib}=A*m_{POSS}+B*{\textrm{color}}+C,
\end{equation}
and $m=O,E,J,F,N$ ($m_{SDSS}$ are defined by eq. \ref{synthetic}). This step removes 
systematic magnitude errors due to local non-linearities of the plate, color-term dependence 
and zero-point offsets\footnote{A similar procedure was used by Munn et al. (2004) 
for astrometric recalibration of POSS catalogs.}. In the second step, we use all the ``good'' 
sources from a given Schmidt plate ($\sim$36 deg$^{2}$) to correct the dependence of the 
$m_{recalib}-m_{SDSS}$ residuals on magnitude (using median $m_{recalib}-m_{SDSS}$ in 
1 mag wide bins, and linear interpolation between the bin centers). Such residuals are 
typically larger at the faint end, and are probably caused by incorrect sky estimates in 
the POSS catalogs.

\subsubsection{              The Optimal Recalibration Scale                  }

\begin{figure}
\plotfiddle{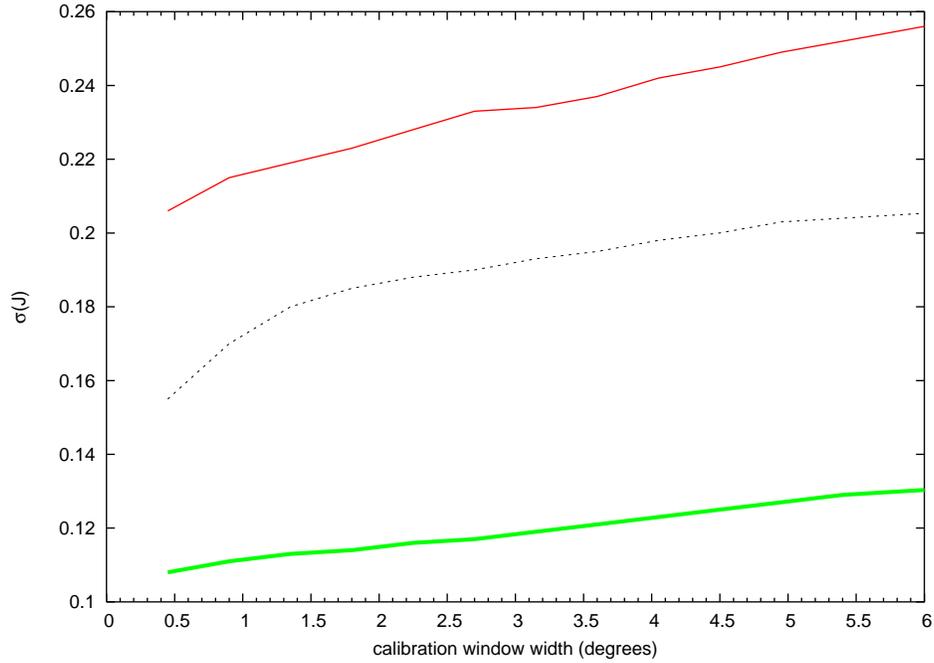}{6cm}{270}{50}{50}{-200}{320}
\caption{
Photometric accuracy as a function of the recalibration window width for three POSS-II $J$ plates.
The decrease of the window width decreases photometric errors as expected. Note
that when extrapolating curves to zero window width, plates show varying photometric accuracy, 
reflecting different intrinsic properties. The estimated improvement in accuracy by decreasing 
the window width from $\sim$0.5$^\circ$ to 0 is only $\sim$0.01-0.02 mag. 
\label{scale}
}
\end{figure}

As advocated by Lattanzi \& Bucciarelli (1991), the characteristic scale for inhomogeneities
in Schmidt plates is about 0.5$^\circ$. We tested their result by recalibrating POSS-II $J$ band 
plates by varying the calibration window width. Fig.~\ref{scale} shows the final errors as a
function of that width for three plates. As expected, the decrease of calibration window width 
decreases photometric errors all the way to practical limit of $\sim$0.5$^\circ$ set by the
minimum number of required calibration stars. The figure demonstrates that improvement in accuracy 
by decreasing the window width from $\sim$0.5-1$^\circ$ to 0 is only $\sim$0.01 mag, thus
confirming the result of Lattanzi \& Bucciarelli. Note that when extrapolating curves to zero 
window width, plates show varying photometric accuracy, reflecting different intrinsic properties. 

\begin{figure}
\plotfiddle{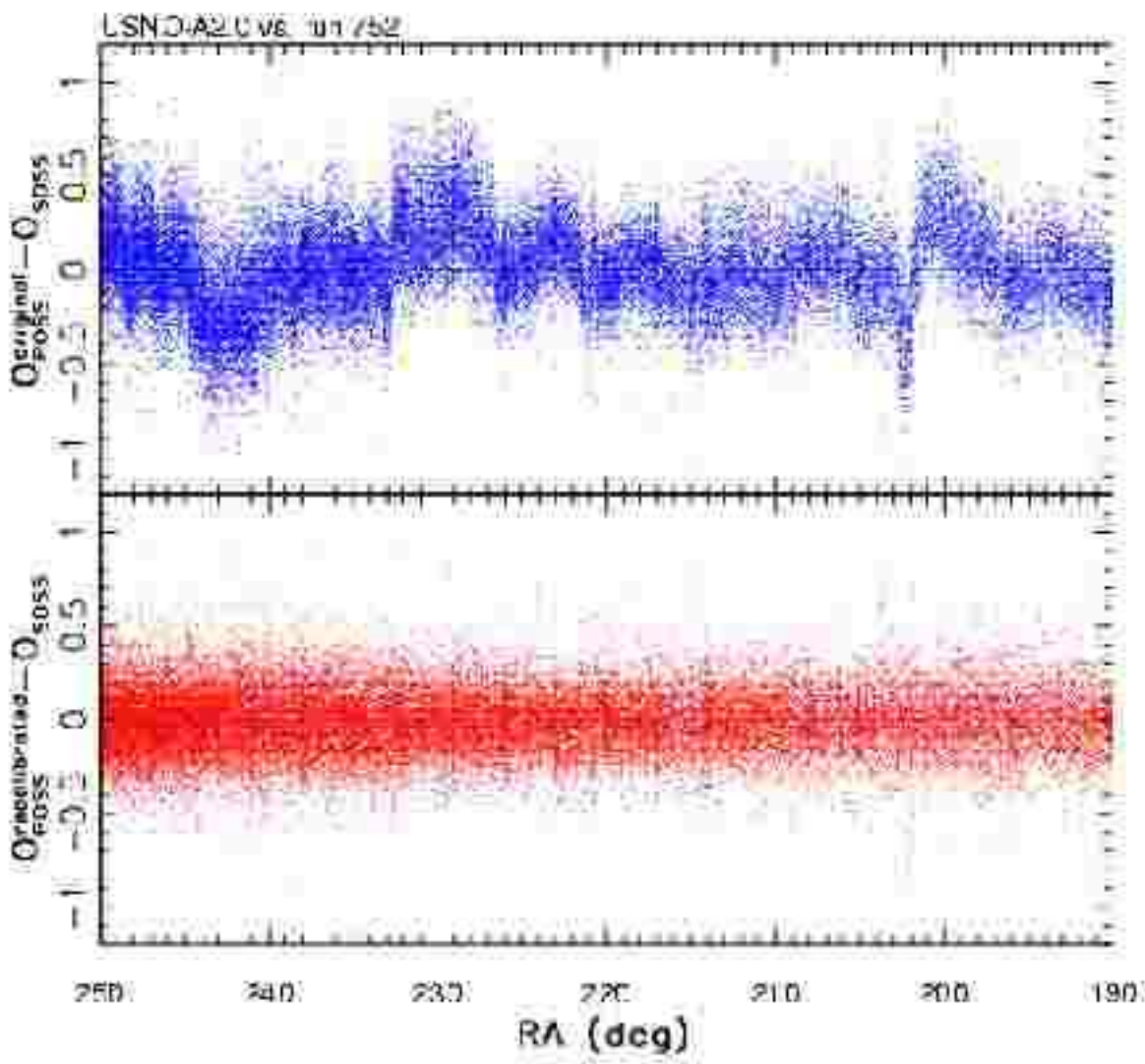}{14cm}{0}{85}{85}{-260}{-220}
\caption{
An illustration of the systematic photometric errors in POSS I catalogs,
and of improvements made possible thanks to a dense grid of calibration
stars provided by SDSS. The top panel shows the differences between the 
original USNO-A2.0 $O$ magnitudes and synthetic SDSS-based $O$ magnitudes
for isolated stars with $O_{SDSS}<18.5$, from a narrow equatorial strip 
(SDSS run 752, $|Dec|<1.25$). Note the large jumps at the boundaries of 
6 degree wide Schmidt plates. The bottom panel shows the differences between 
the recalibrated USNO-A2.0 $O$ magnitudes and synthetic SDSS-based $O$ 
magnitudes for the same stars. 
\label{dmvsRAUSNO}
}
\end{figure}

\begin{figure}
\plotfiddle{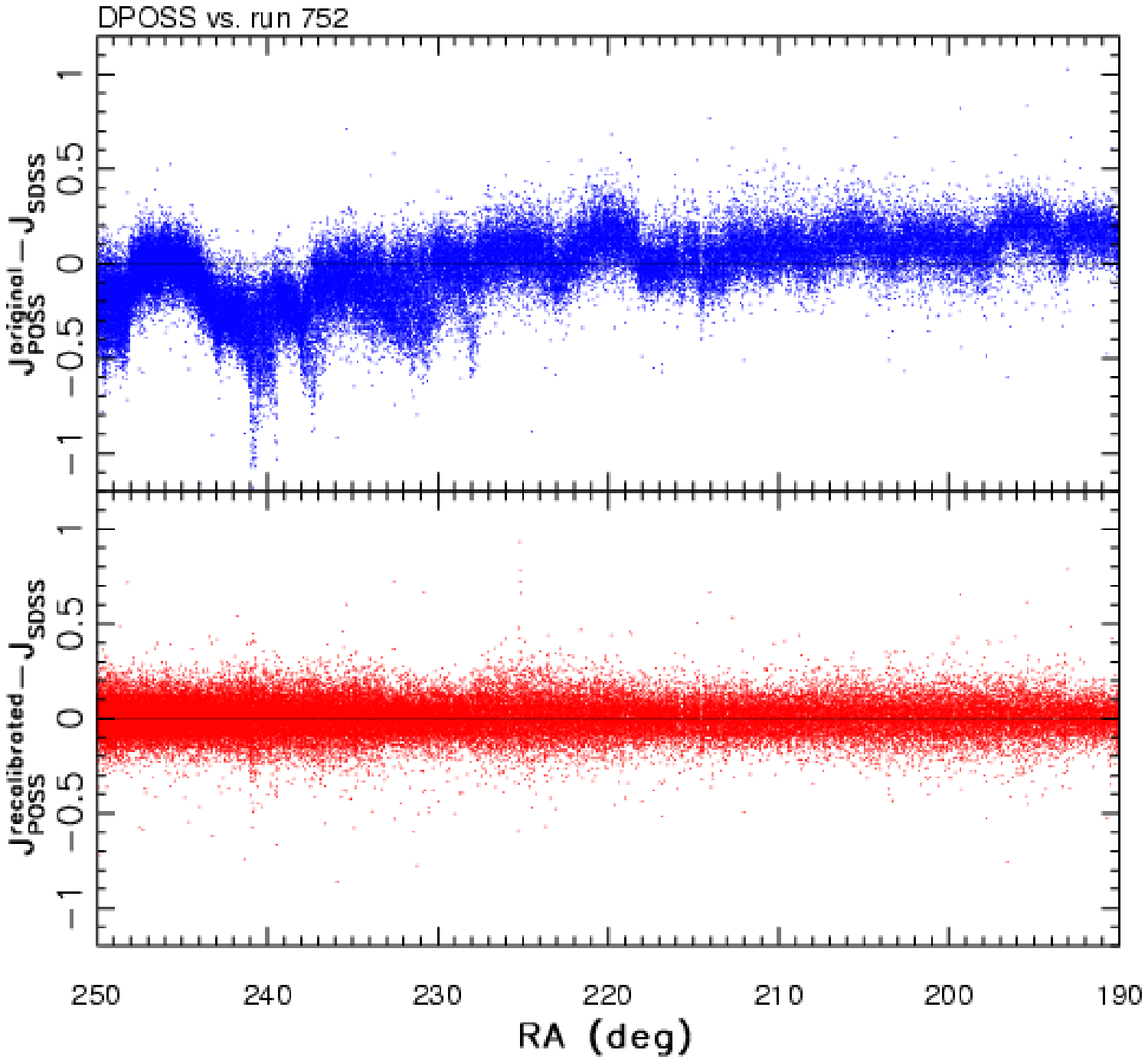}{14cm}{0}{85}{85}{-260}{-220}
\caption{
Same as Fig.~\ref{dmvsRAUSNO}, except that the G magnitudes taken from 
a POSS II-based DPOSS catalog are shown.
\label{dmvsRAdposs}
}
\end{figure}

\begin{figure}
\plotfiddle{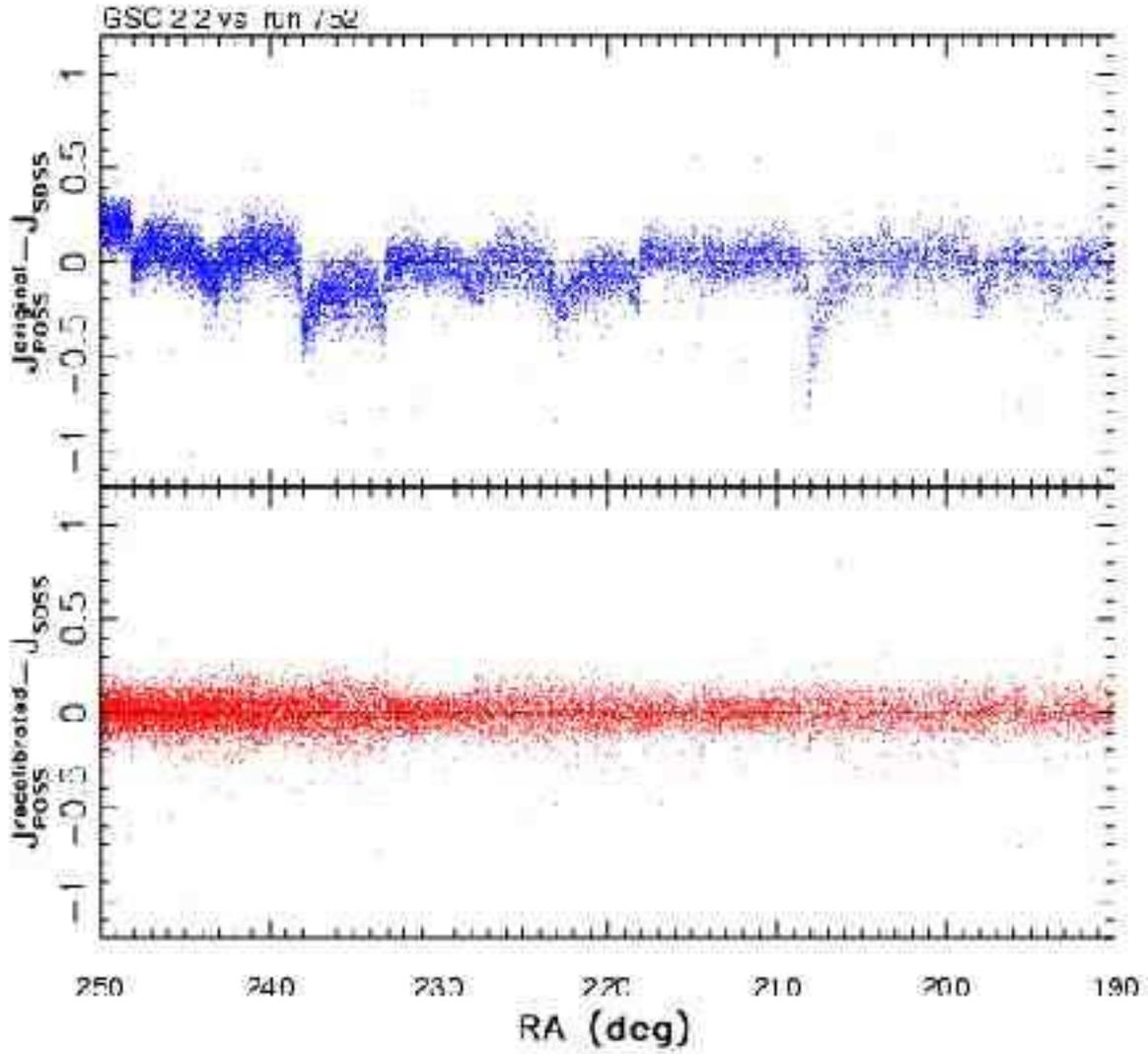}{14cm}{0}{85}{85}{-260}{-220}
\caption{
Same as Fig.~\ref{dmvsRAUSNO}, except that the J magnitudes taken from 
a POSS II-based GSC2.2 catalog are shown.
\label{dmvsRAgsc}
}
\end{figure}

The plate-dependent systematic photometric errors in POSS catalogs are illustrated 
in Figs.~\ref{dmvsRAUSNO}--\ref{dmvsRAgsc} (the behavior for USNO-B1.0 catalog is similar). 
The large jumps in photometric
errors at the boundaries of 6 degree wide Schmidt plates are obvious, and 
suggest that the photometric recalibration of POSS data is mandatory when
searching for variable sources that vary less than a few tenths of a magnitude.

\subsection{              Analysis of the Recalibration Results               }

\subsubsection{         Recalibration Results for the USNO Catalogs           }

\begin{figure}
\epsscale{1}
\plotone{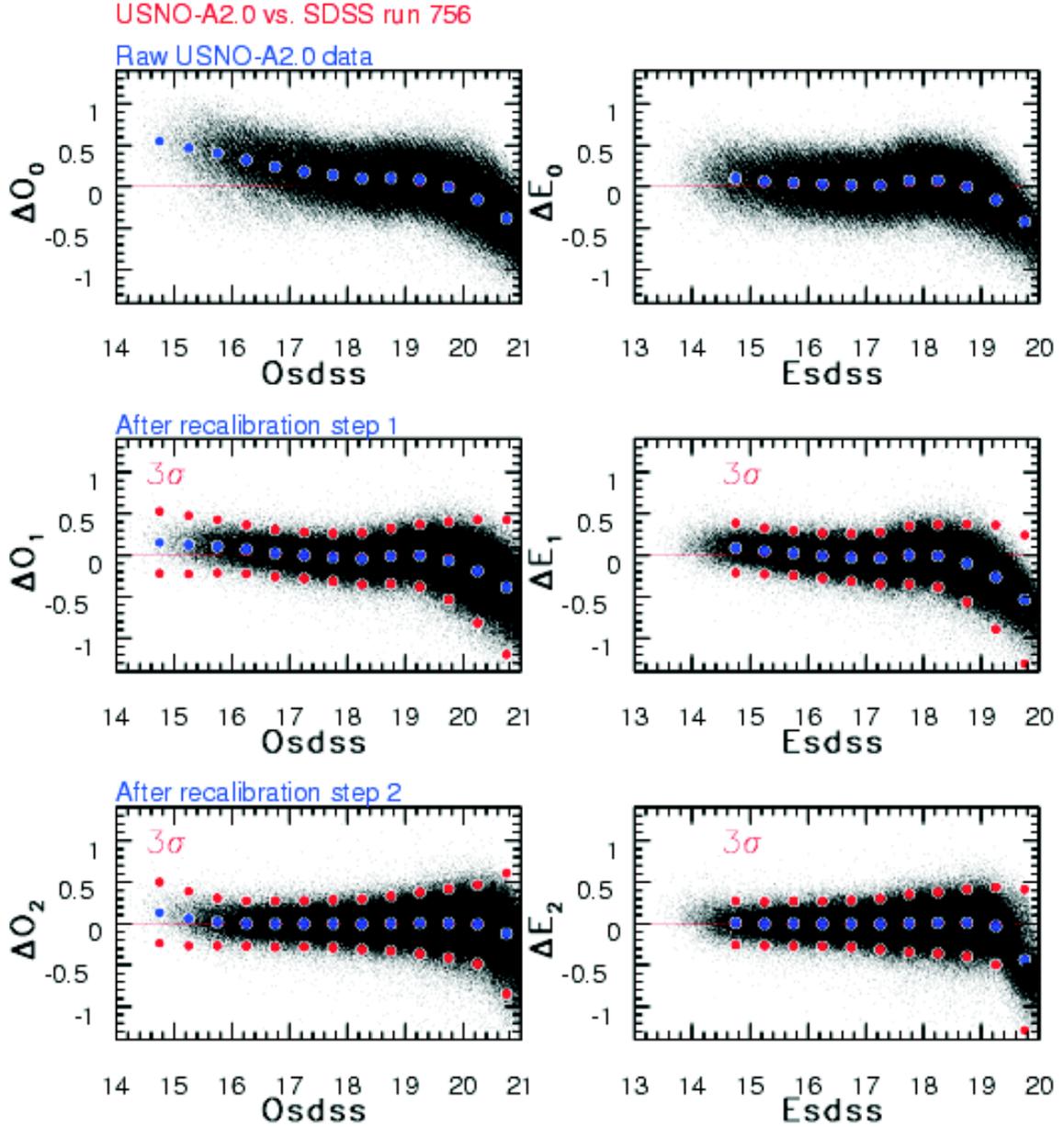}
\caption{
The illustration of the recalibration method for the USNO-A2.0 catalog. The dots in the top panels
represent magnitude differences between the original POSS $O$ (left column) and $E$ (right column) 
magnitudes and the synthetic SDSS-based $O$ and $E$ magnitudes, as a function of the latter,
for about 300,000 stars observed in $\sim$100 deg$^{2}$ of sky in SDSS run 752. The middle panel 
shows the magnitude differences after the first recalibration step, where color-term and zero-point
systematic errors are removed. The results of the second recalibration step, which removes the
dependence of magnitude differences on magnitude, are shown in the bottom panels.  The middle set
of large symbols in each panel shows the median differences in magnitude bins, and the two
outer sets of large symbols show  equivalent Gaussian widths (determined from the interquartile range), 
multiplied by 3.
\label{stepsA}
}
\end{figure}

The results of two recalibration steps, described in the previous section, are illustrated 
for the USNO-A2.0 catalog in Fig.~\ref{stepsA}. As evident from the middle panels,
the first step results in smaller scatter between SDSS and recalibrated POSS magnitudes, 
but the magnitude dependence of their differences remains appreciable. This dependence
is removed in the second recalibration step, as discernible from the bottom panels. 

\begin{figure}
\plotfiddle{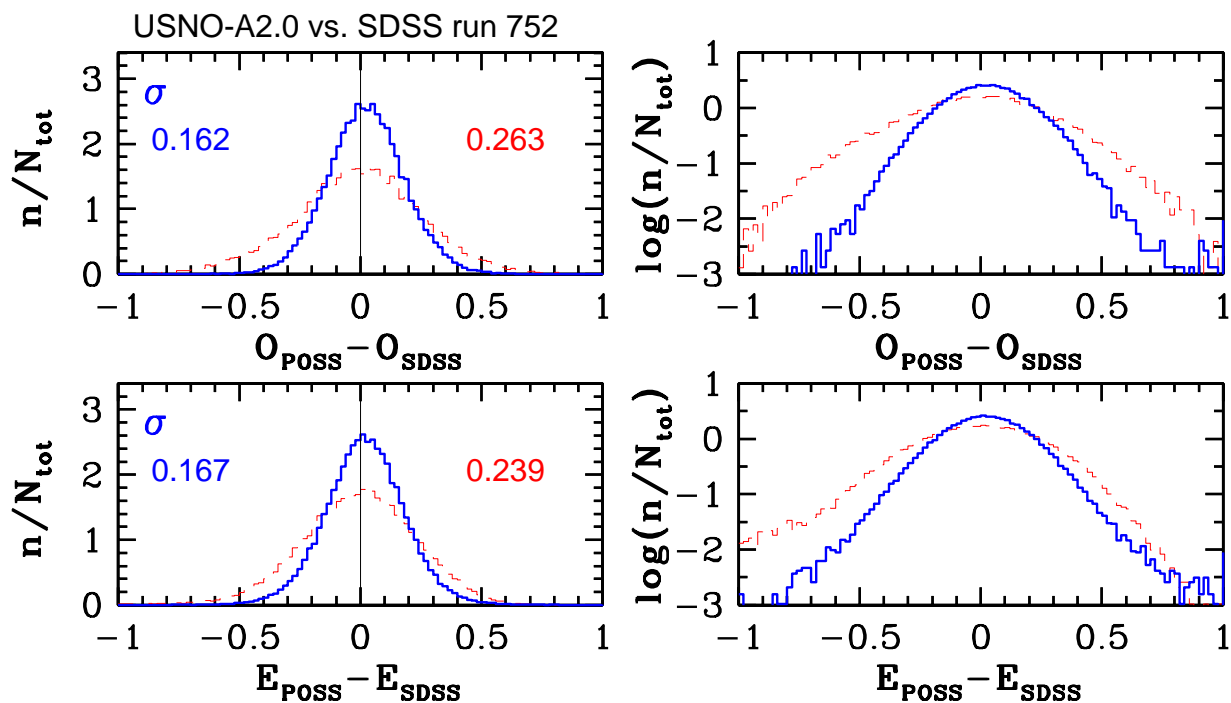}{8.5cm}{0}{90}{90}{-260}{-400}
\caption{
The improvements in photometric errors after recalibration for the USNO-A2.0 catalog.
The POSS-SDSS magnitude differences before recalibration are shown by thin lines, and 
those after recalibration by thick lines. The left column shows error distribution 
on a linear scale, and the right column on a logarithmic scale. The equivalent Gaussian 
widths, determined from the interquartile range, are also shown in each panel 
(right: before, left: after).
\label{tailsA}
}
\end{figure}

The recalibration procedure generally results in about a factor of 2 improvement in the 
root-mean-square (rms) scatter between SDSS-based synthetic POSS magnitudes and the measured 
POSS magnitudes. Fig.~\ref{tailsA} compares the POSS-SDSS magnitude differences before (thin 
lines) and after (thick lines) calibration, for $O$ and $E$ bands, on a linear and logarithmic
scale. As evident, the recalibration not only results in a smaller rms scatter, but also
significantly clips the tails. Both effects are of crucial importance when selecting variable
objects.  

\begin{figure}
\plotfiddle{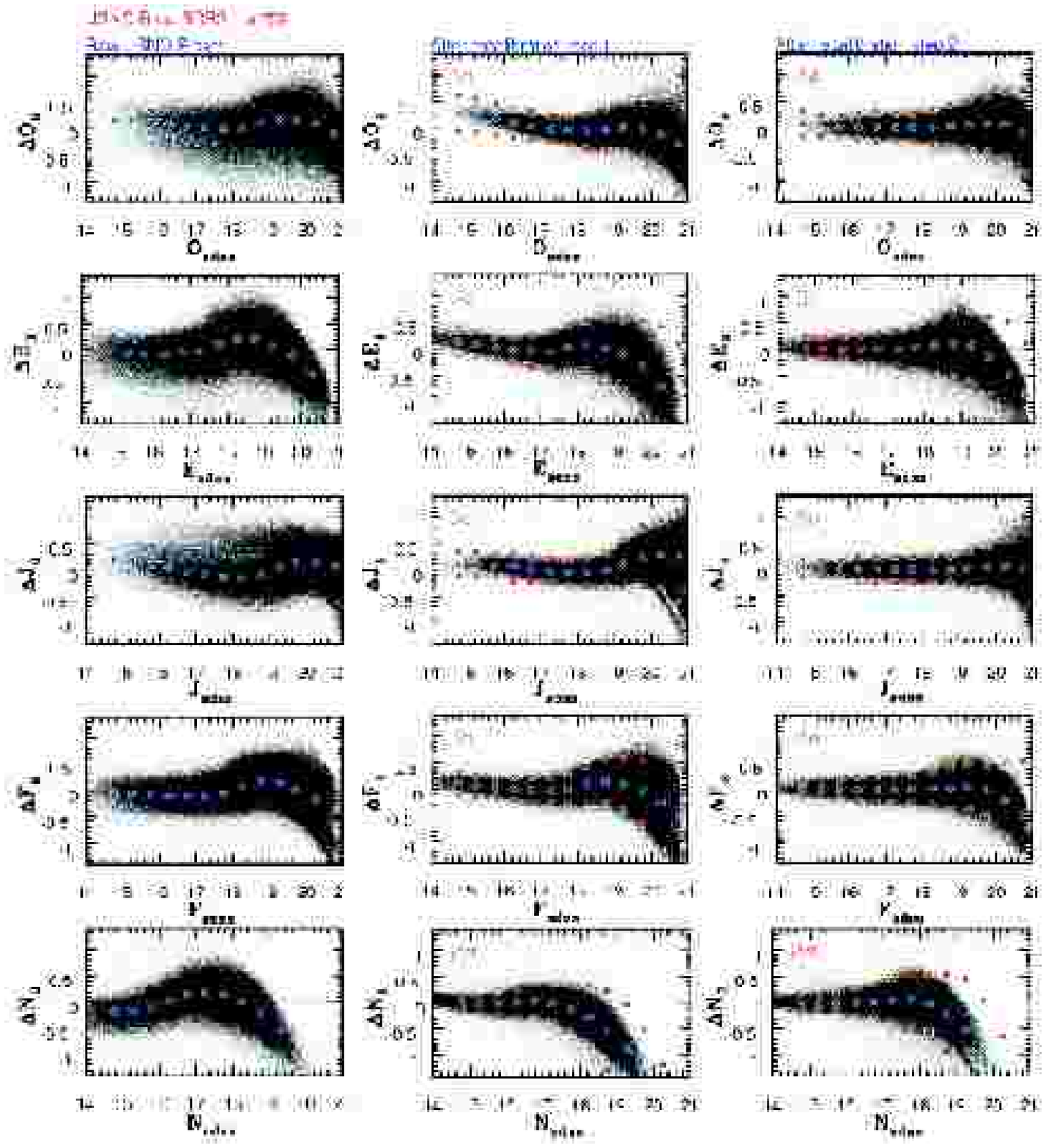}{18cm}{0}{90}{90}{-250}{-30}
\caption{Same as Fig.~\ref{stepsA}, except for the USNO-B1.0 catalog.
\label{stepsB}
}
\end{figure}

\begin{figure}
\plotfiddle{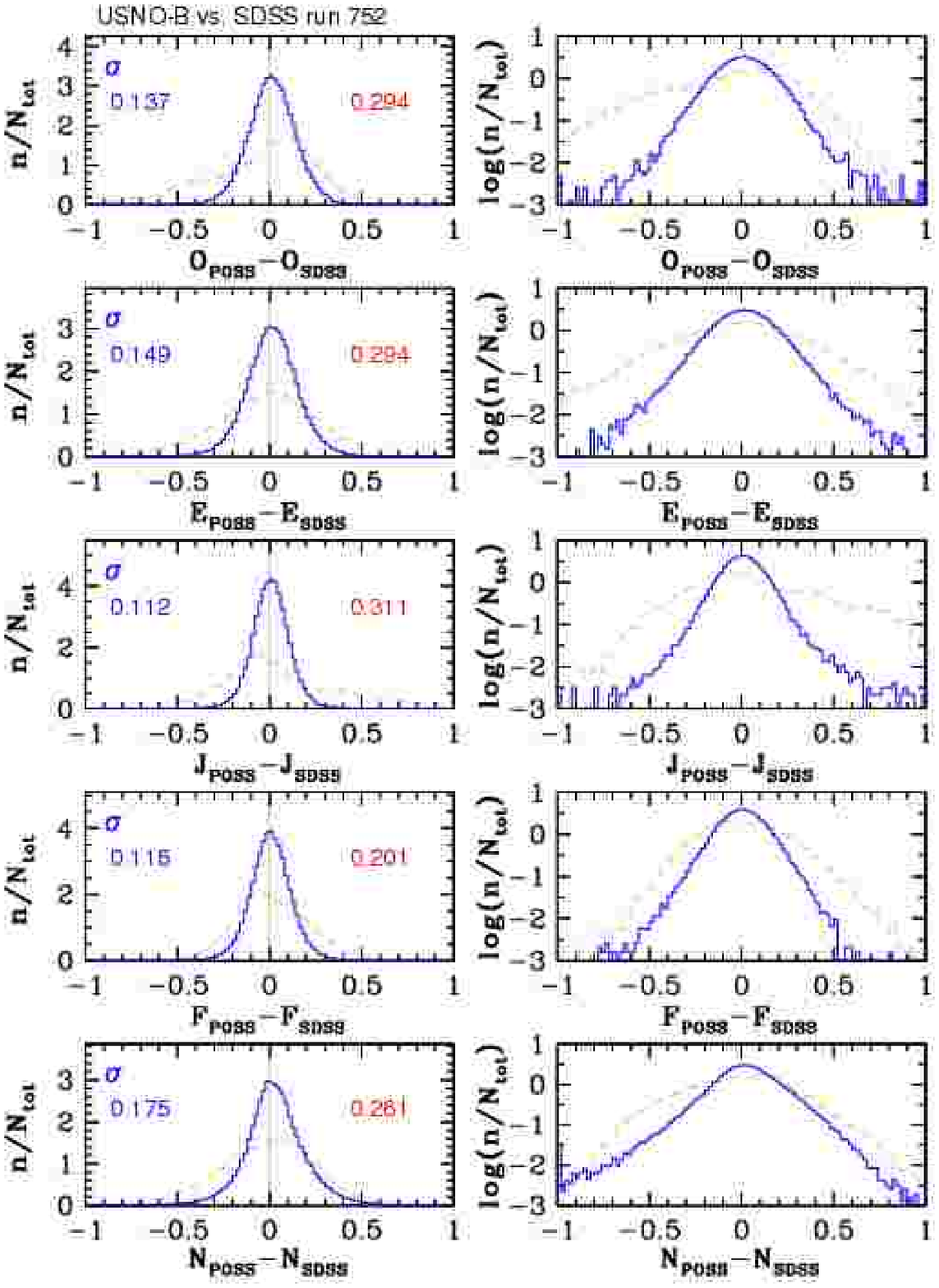}{19cm}{0}{80}{80}{-230}{-40}
\caption{
Same as Fig.~\ref{tailsA}, except for the USNO-B1.0 catalog.
\label{tailsB}
}
\end{figure}

The corresponding results for the USNO-B1.0 catalog are shown in Figs.~\ref{stepsB} and 
\ref{tailsB}. We note that the original \O\ and \E\ magnitudes have somewhat smaller errors
in the USNO-A2.0 catalog, but the USNO-B1.0 magnitudes are marginally better after recalibration.

\subsubsection{        Recalibration Results for the GSC2.2 and DPOSS Catalogs        }

\begin{figure}
\plotfiddle{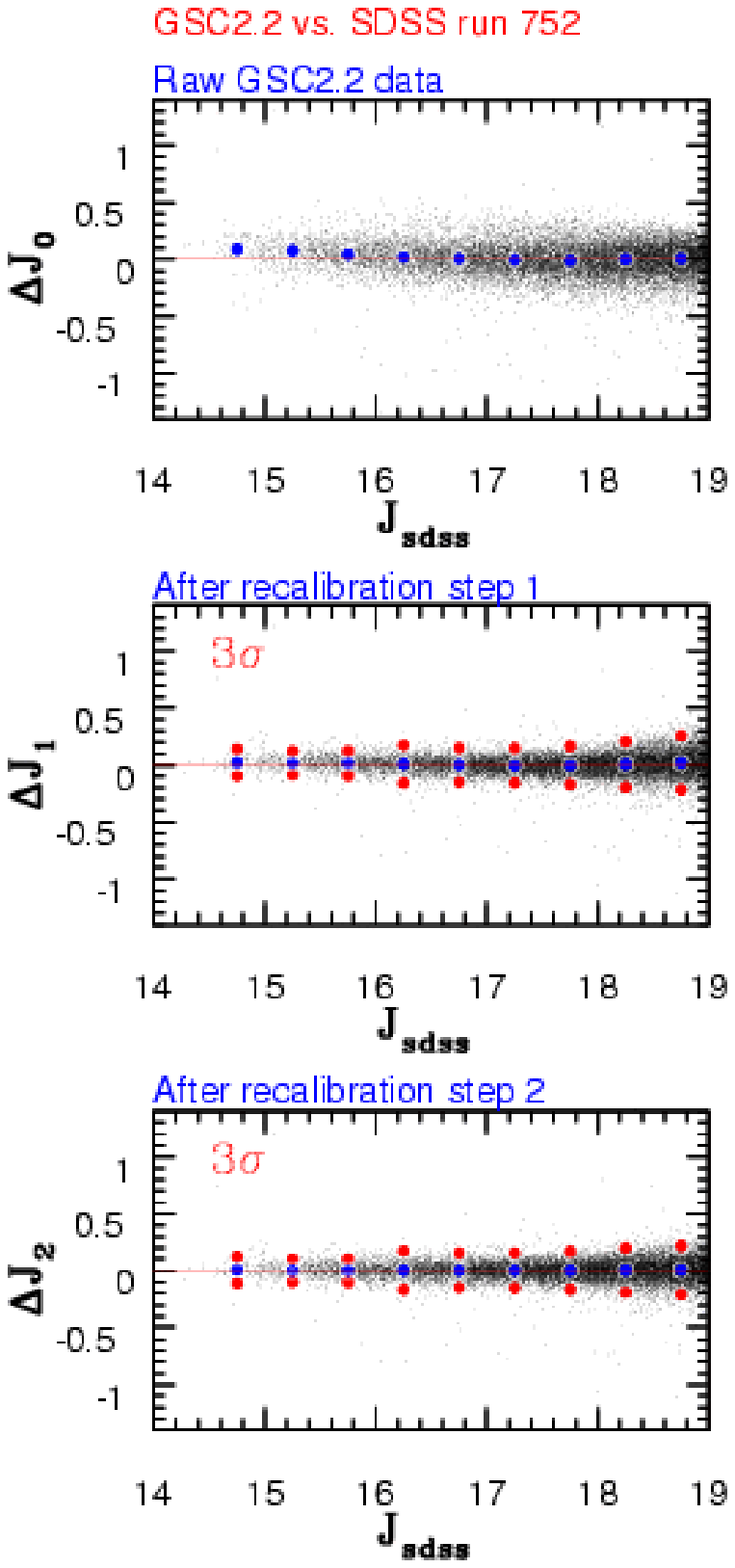}{6cm}{0}{80}{80}{-250}{-280}
\plotfiddle{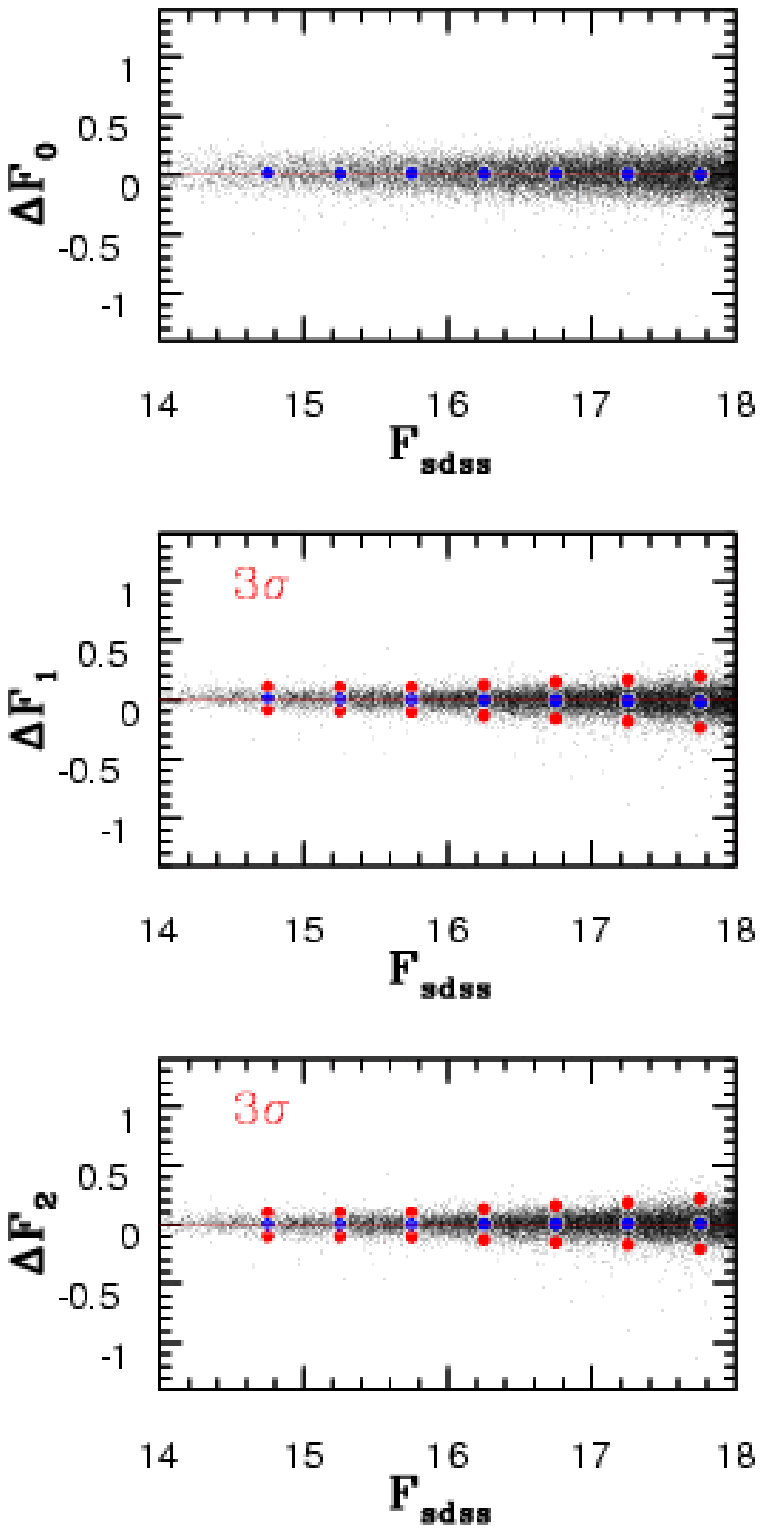}{6cm}{0}{80}{80}{-20}{-92}
\caption{
Same as Fig.~\ref{stepsA}, except that only the first recalibration step is shown, 
for the GSC2.2 catalog.
\label{stepsG}
}
\end{figure}

\begin{figure}
\plotfiddle{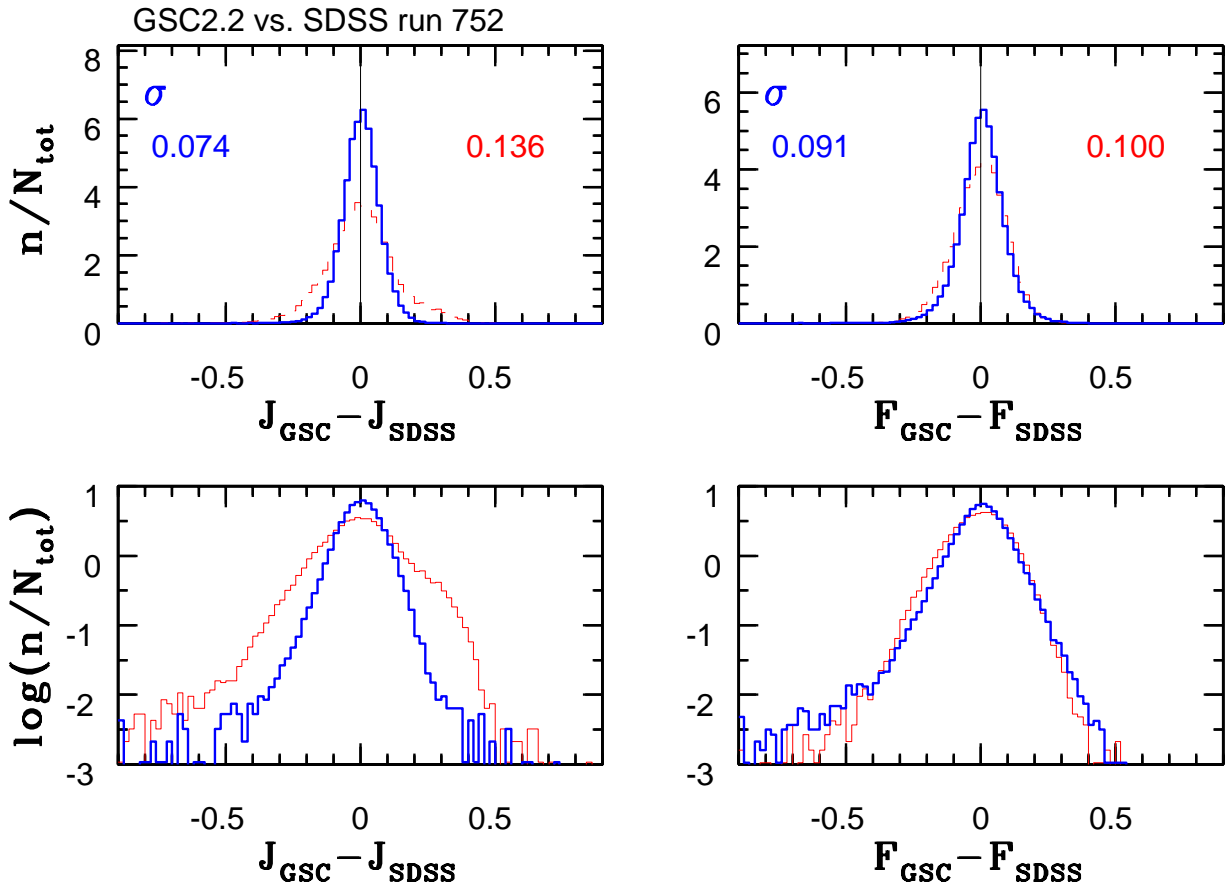}{11cm}{0}{90}{90}{-260}{-290}
\caption{
Same as Fig.~\ref{tailsA}, except for the GSC2.2 catalog.
\label{tailsG}
}
\end{figure}

\begin{figure}
\plotfiddle{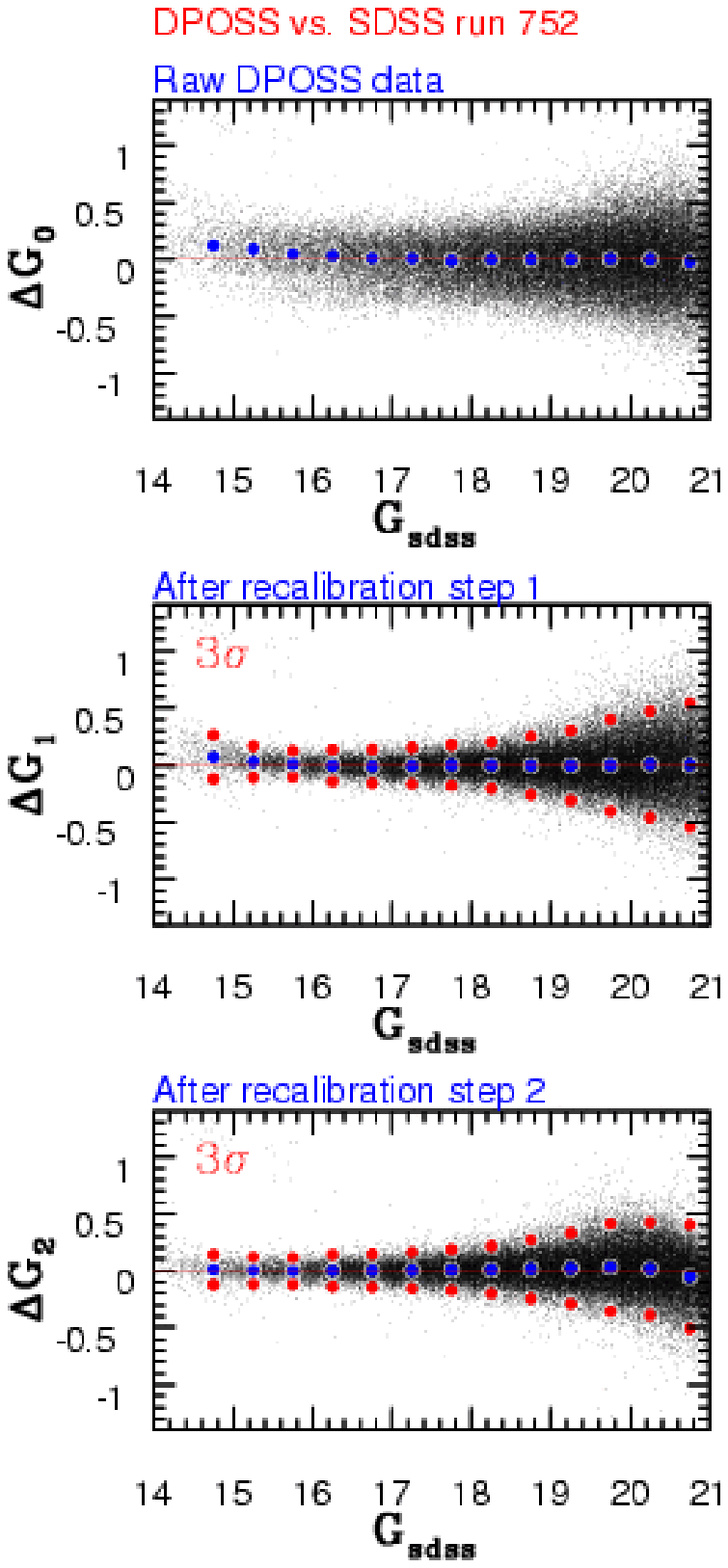}{6cm}{0}{80}{80}{-250}{-280}
\plotfiddle{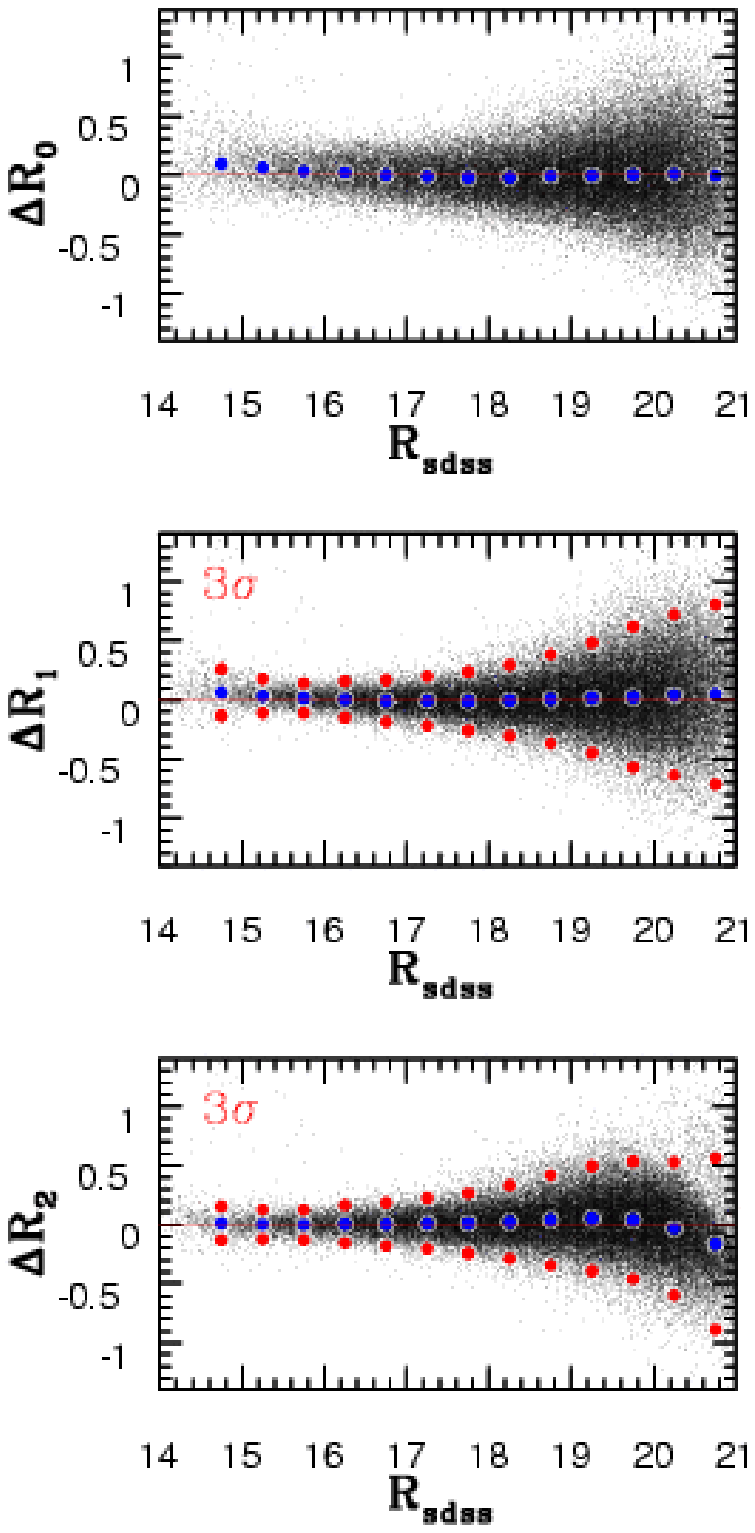}{6cm}{0}{80}{80}{-20}{-92}
\caption{
Same as Fig.~\ref{stepsA}, except for the DPOSS catalog.
\label{stepsD}
}
\end{figure}

\begin{figure}
\plotfiddle{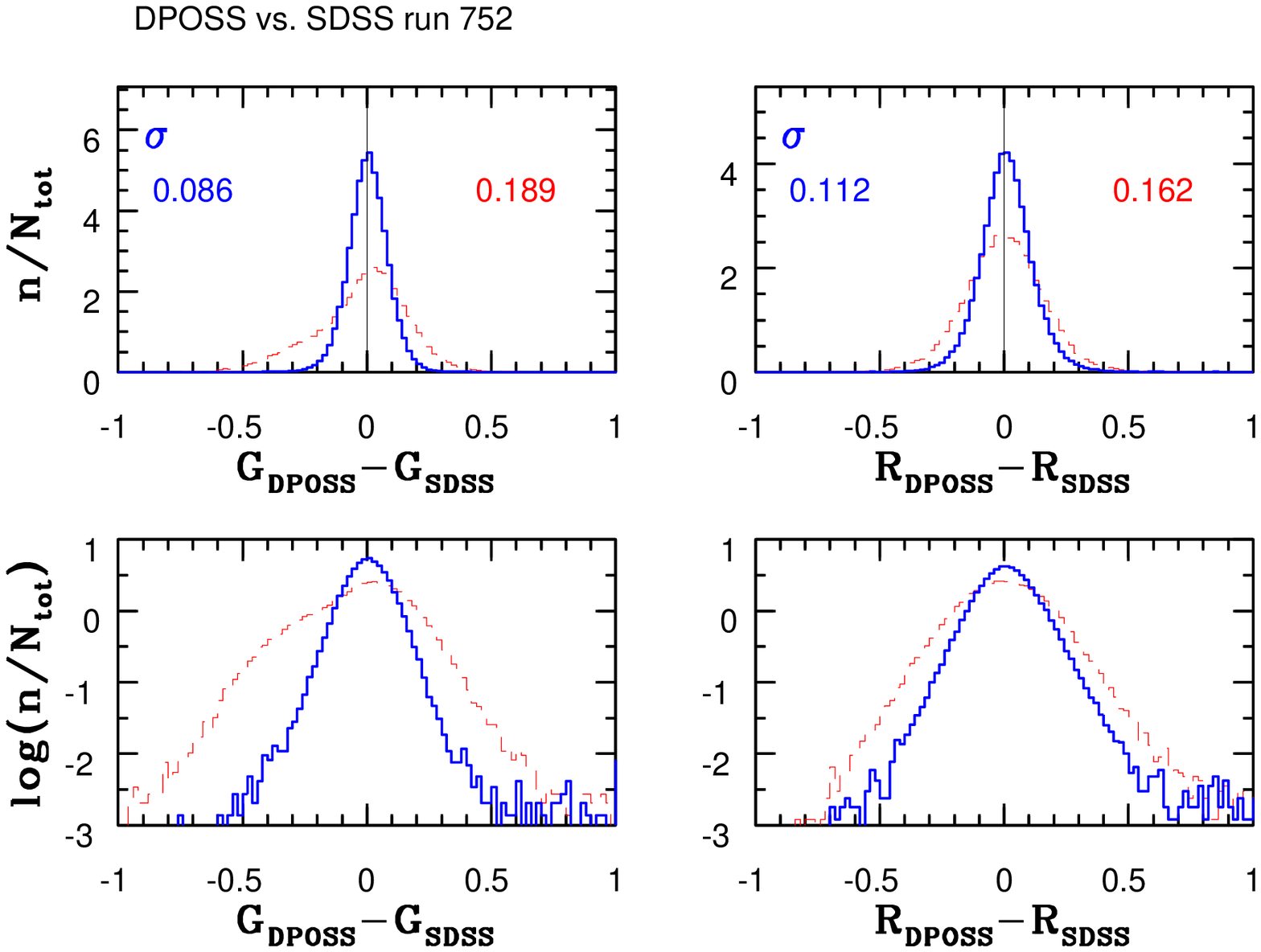}{11cm}{0}{90}{90}{-260}{-270}
\caption{
Same as Fig.~\ref{tailsA}, except for the DPOSS catalog.
\label{tailsD}
}
\end{figure}

The main difference between the recalibration of GSC2.2 and other catalogs is the lack
of plate information\footnote{Observations listed in the catalog are collected
from multiple plates, even if confined to a small sky region, and the plate number
from which a particular entry was derived is not provided.} which prevented the 
second recalibration step. Nevertheless, 
Figs.~\ref{stepsG} and \ref{tailsG} show that for the GSC2.2 catalog the magnitude
dependence of the differences between SDSS and recalibrated POSS magnitudes is minor.
The results for the DPOSS catalog are shown in Figs.~\ref{stepsD} and \ref{tailsD}.

\subsubsection{      Summary of  Recalibration Results    }

\begin{figure}
\plotfiddle{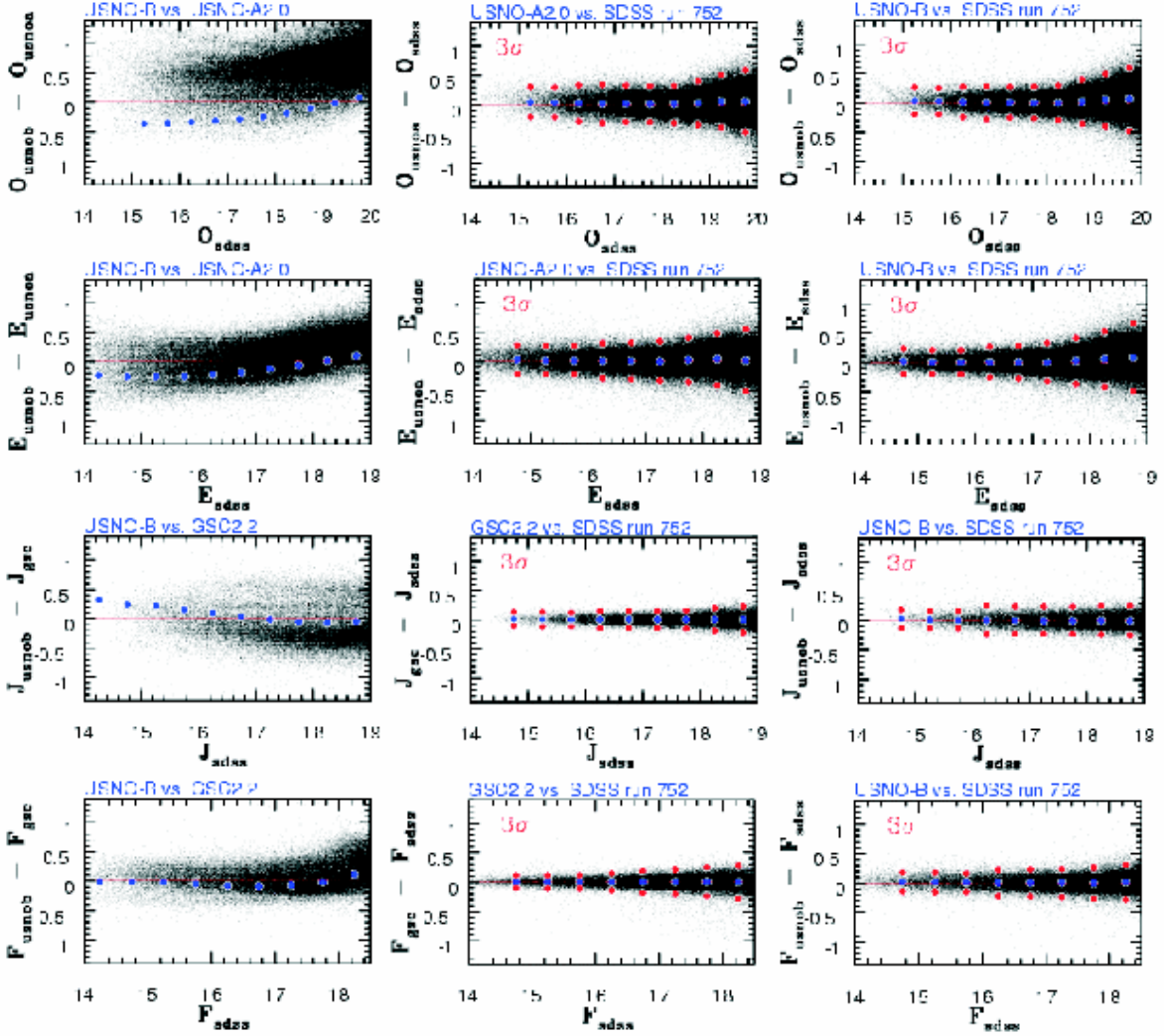}{11cm}{0}{95}{95}{-290}{-180}
\caption{
A summary of recalibration results for different POSS catalogs, as marked. The POSS-SDSS
comparisons, shown in the middle and right columns, are based on the recalibrated magnitudes, 
while the POSS-POSS comparisons for different input catalogs (left column) are based on their 
original magnitudes.
\label{comparison}
}
\end{figure}

A summary comparison of the original and recalibrated magnitudes is shown in 
Fig.\ref{comparison}. The rms values of magnitude differences before and after the recalibration 
for all the catalogs and bands are listed in Table 2. The final errors for the POSS II 
magnitudes are generally smaller ($\sim$0.10 mag) than for the POSS I magnitudes 
($\sim$0.15 mag), both evaluated for stars brighter than $g=19$. The smallest final errors 
are obtained with the GSC2.2 catalog for which they approach 0.07 mag at the bright end.

\section{     Preliminary Analysis of the POSS-SDSS Catalogs of Variable Sources           } 

The comparison of the SDSS photometric catalog with photometrically recalibrated POSS catalogs
can yield a large number of variable sources. Various methods can be employed to produce
such a list of candidate variables, depending on whether each band/catalog is considered
separately or not, on the cutoff values for magnitude differences, the sample faint
limit, etc. The DPOSS catalog is the main catalog used in the subsequent analysis (while 
the smallest final errors are produced with the GSC2.2 catalog, its public version is not 
as deep as the DPOSS catalog). For the POSS I survey we chose the USNO-A2.0 catalog (the other
option is USNO-B1.0; the photometric errors after recalibration are similar for both USNO catalogs) 
because it is distributed as a part of the SDSS Data Release 1 (Abazajian et al. 2003). 

The criteria for selecting candidate variable sources are described in the next section,
and a series of tests for estimating the selection reliability are described in the
subsequent section. 

\subsection{                         Selection Criteria                    }

When selecting candidate variable sources we consider each POSS band individually for 
two reasons. First, sometimes POSS observations of the same sky regions were not obtained
at the same time, and treating each epoch separately increases the selection 
completeness for sources variable on short time scales. Second, additional constraints
that combine different bands (e.g. ``a source must vary in both $O$ and $E$ band'') 
can be easily imposed after the initial single-band based selection. We consider 
only isolated point sources detected by both POSS and SDSS. An initial attempt to 
isolate orphaned sources detected by only one survey was impeded by the large number 
of false positives due to problems with POSS plates, and is postponed for future
analysis.

For each catalog and band we define the faint magnitude limit, $m_{faint}$, minimum flux 
variation, $\Delta m = |m_{SDSS}-m_{POSS}|$, and its minimum significance, 
$\chi = \Delta m/\sigma$.  For the photometric error, $\sigma$, we take the
rms scatter for all the stars in a given calibration patch and in 0.5 mag wide 
magnitude bins (using a linear interpolation of binned errors as a function of 
magnitude). The adopted values of selection parameters for each catalog are listed
 in Table 3, as well as the
number of selected candidate variable sources. In general, the $\Delta m$ condition
controls the selection at the bright end, while the $\chi$ condition limits the
sample at the faint end. Thus, the selection efficiency is roughly independent of 
magnitude until about 1-1.5 mag above the adopted faint limit, $m_{faint}$, when 
it starts decreasing. We find that typically 15-20\% of selected candidates
 simultaneously satisfy conditions in two bands from a given catalog.

These particular selection criteria were adopted after a trial and error procedure 
which utilized tests described in the next section. We chose to err on the conservative
side and increase catalog robustness at the expense of its completeness, since the small 
number of epochs already introduced substantial incompleteness. Hence,
the fraction of variable sources reported here is only a lower limit.

\subsection{         Tests of the Selection Reliability     }

Given the selection criteria described in the previous section, it is 
necessary for subsequent analysis to estimate the completeness and 
efficiency of the resulting samples. The selection completeness, 
the fraction of true variable sources in the analyzed sky region and
observed magnitude range selected by the algorithm, is certainly low 
because the selection is based on variations in only one bandpass, and
a fairly large $\Delta m$ cutoff compared to the typical amplitudes of 
variable sources (e.g. most RR Lyrae stars and quasars have peak-to-peak 
amplitudes $\la$1 mag). For example, Ivezi\'{c} et al. (2000) used 
two-epoch SDSS measurements to select candidate RR Lyrae stars and obtained 
completeness of $\sim$50\% for a $\Delta m$ cutoff of 0.15 mag. With 
larger $\Delta m$ cutoffs adopted here, the expected selection completeness 
for RR Lyrae stars is about 20\% (see Section~\ref{RRLyr} for a 
direct measurement). The completeness for other types of variable source
depends on the shape and amplitudes of their light curves and is hard to 
estimate, but for most sources is similarly low. While such a low completeness 
cannot be avoided with the available data, its stability across the sky
can be controlled. Such a stability is demonstrated by the lack of features 
in  the distribution of quasars selected by variability, as well as by 
the recovery of known structures in the distribution of RR Lyrae stars,
as discussed below. 

The selection efficiency, the fraction of true variable sources in 
the selected sample, may severely impact the analysis if not
sufficiently large. We demonstrate using a series of tests that 
the selection efficiency is indeed very large (73\%) and thus allows 
a robust analysis of variable faint optical sources.

The main diagnostic for the robustness of the adopted selection criteria is the
distribution of selected candidates in SDSS color-magnitude and color-color
space. Were the selection a random process, the selected candidates would
have the same distribution as the whole sample. However, we find that the samples
of candidate variables have a significantly different distribution, as detailed below. 
The most robust and quantitative test for estimating selection reliability is a comparison 
with repeated SDSS imaging observations, though applicable to only a small
fraction ($\sim$10\%) of the sky area discussed here where such SDSS observations 
exist. Another powerful test for candidates with large suspected flux variation 
($\ga$ 0.5 mag) is a simple visual comparison of POSS and SDSS images. 
While we found a number of spurious candidates using this method, their fraction 
is not large enough to significantly affect our results.

\subsubsection{ The Distribution of Candidate Variable Sources in SDSS Color-color Diagrams  }

\begin{figure}
\plotfiddle{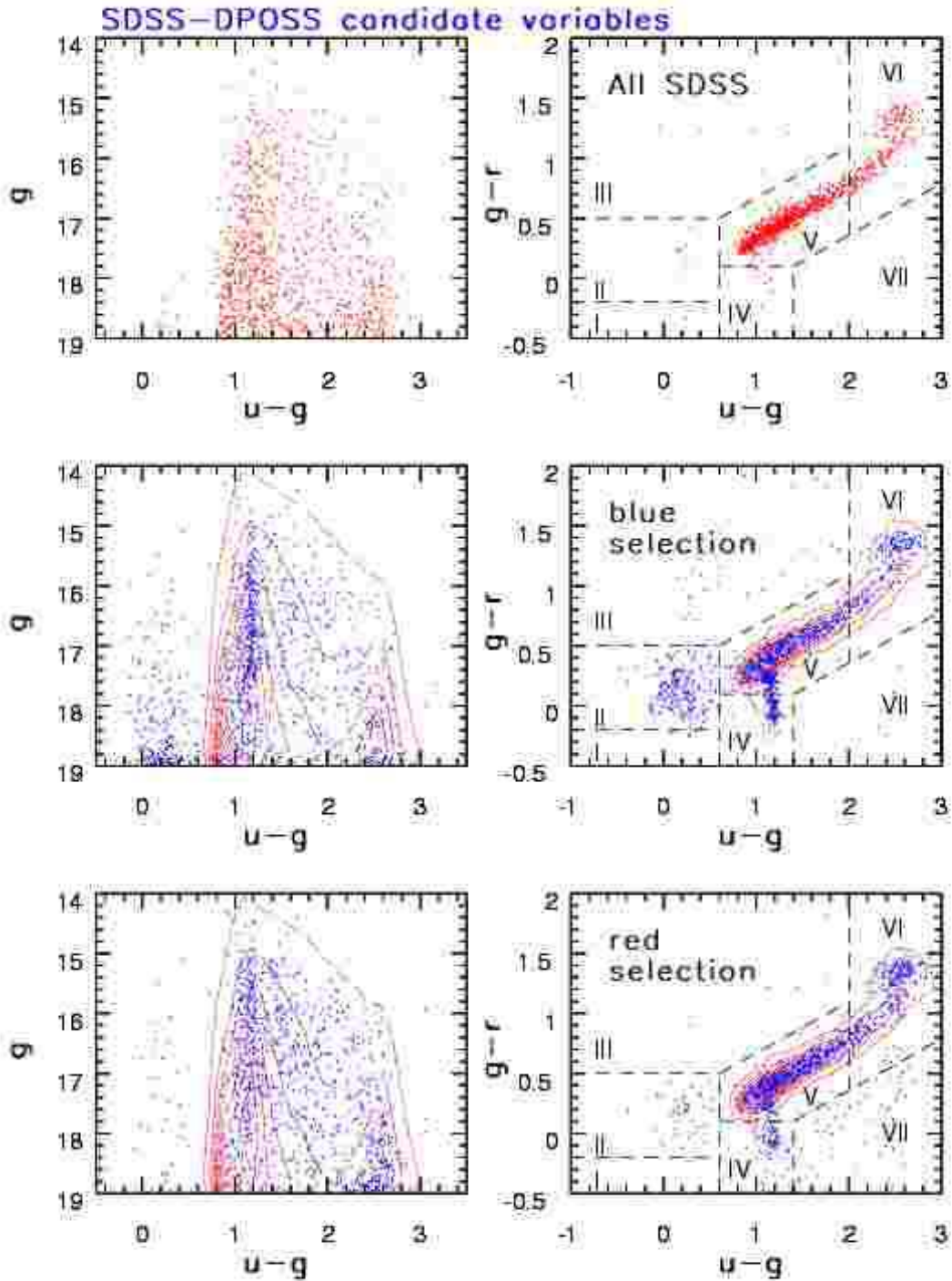 }{17.5cm}{0}{78}{78}{-230}{-60}
\caption{
The distribution of SDSS-DPOSS candidate variable sources with $g<19$
in representative SDSS color-magnitude (left) and color-color diagrams 
(right). The top row is shown for reference and displays a sample 
of SDSS point sources with the same flux limit and with the same total number
of sources ($\sim$1000). The middle and bottom rows display the distributions 
from the top row by contours, and variable sources selected from the DPOSS 
catalog as dots ($G$ selection in the middle row and $R$ selection in the bottom 
row, see the last two entries in Table 3). The regions marked in the right column 
are used for quantitative comparison of the overall and variable source 
distributions (see Table 4). 
\label{CCDs}
}
\end{figure}

The position of a source in SDSS color-magnitude and color-color diagrams 
is a good proxy for its classification. The distribution of selected 
candidate variables with $g<19$ (using the DPOSS catalog, and restrictive selection
criteria, entries 7 and 8 from Table 3) in representative diagrams is shown in 
Fig.~\ref{CCDs} (the measured magnitudes are corrected for interstellar 
extinction using the map from Schlegel, Finkbeiner \& Davis 1998).
The top row is shown for reference and displays a sample of randomly selected
SDSS point sources with the same flux limits as used for selecting variable 
sources. The middle and bottom rows compare the distributions of this reference
sample, shown as contours, to the distributions of candidate variable sources, 
shown as dots.

The distributions of candidate variable sources and those of the reference sample 
are different, demonstrating that the candidate variables are {\it not} randomly
selected from the whole sample. The most obvious difference between the distributions
is a much higher fraction of quasars (recognized by their UV excess, $u-g < 0.6$)
in the variable sample (quasars are known to be variable on long time scales discussed 
here, see \S \ref{varqso}). Another notable difference is a presence of 
RR Lyrae stars ($u-g \sim 1.2$, $g-r \sim 0$) among the candidate variables.
Thus, known variable sources indeed dominate the selected candidates.

\begin{figure}
\plotfiddle{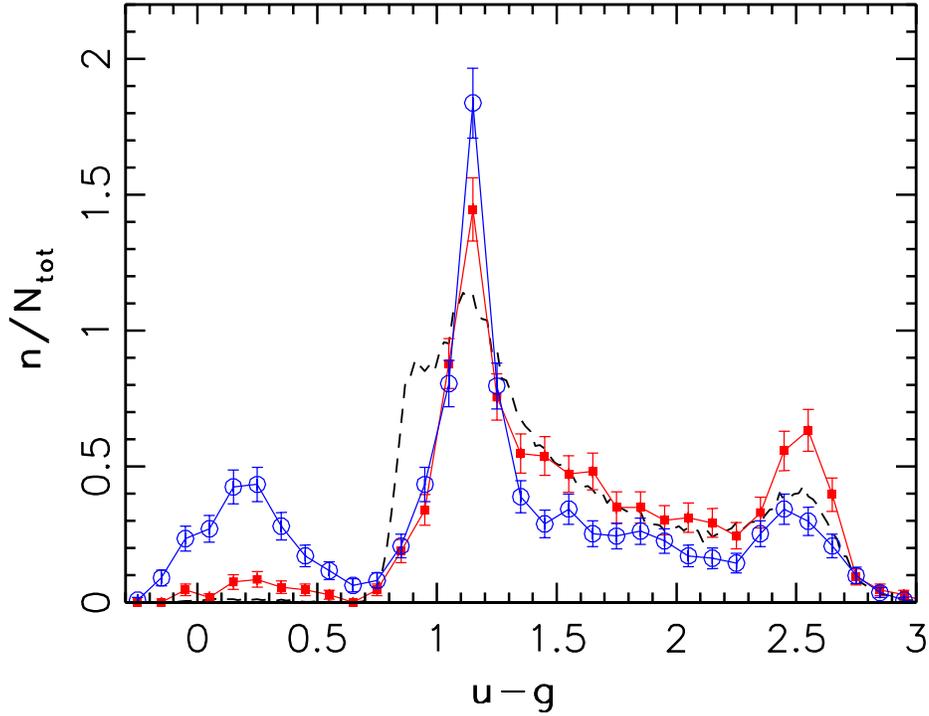}{9cm}{0}{70}{70}{-220}{-240}
\plotfiddle{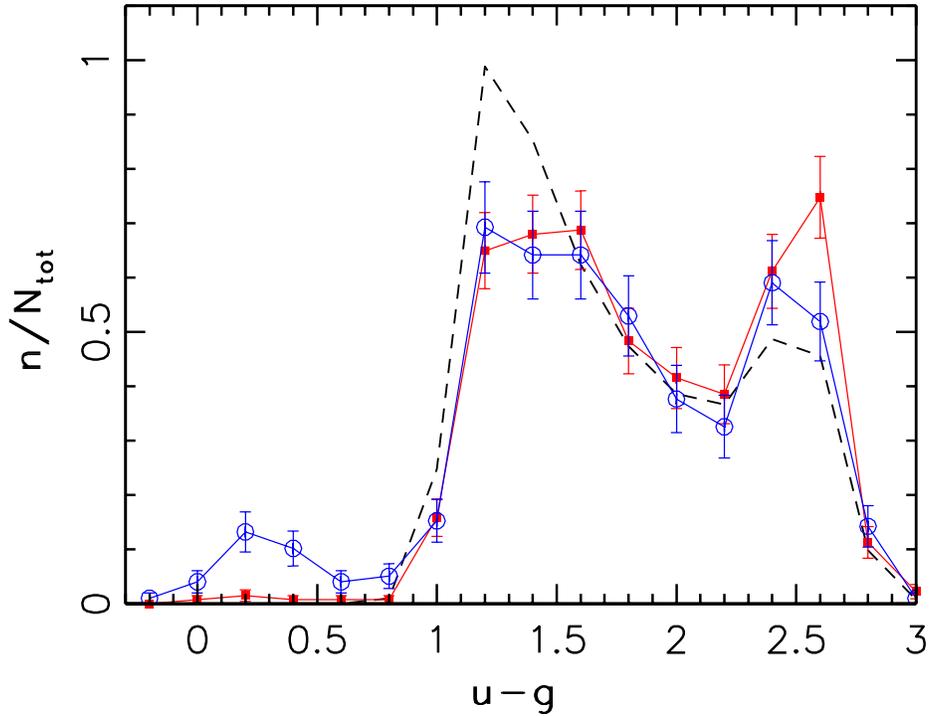}{9cm}{0}{70}{70}{-220}{-260}
\caption{
The comparison of $u-g$ probability density distributions (i.e. the integrals of the 
plotted curves are 1 by definition) for SDSS-DPOSS candidate variable sources 
(symbols with error bars, circles for $G$ selection and squares for $R$ selection),
and for a reference sample with the same magnitude limit (dashed line). The top
panel shows all sources, and the bottom panel shows a subset with $g-r>0.4$ (designed
to avoid the majority of low-redshift quasars, see Fig.~\ref{CCDs}). The peak at 
$u-g\sim0.2$ is dominated by quasars, the peak at $u-g\sim1.15$ by RR Lyrae stars, 
and the peak at  $u-g\sim2.5$ by M stars.
\label{ugHistAll}
}
\end{figure}

In order to quantify these differences, as well as those in other parts of the 
color-color diagram, we divide color-color diagrams into seven characteristic regions, 
each dominated by a particular type of source (for more details about the distribution of 
point sources in SDSS color-color diagrams see Lenz et al. 1998, Fan 1999, Finlator et al. 
2001, and Richards et al. 2002). The fractions of variable and all sources in each 
region are listed in Table 4. Notably, the fraction of variable sources which are 
found in region II, representative of numerous low-redshift quasars, is $\sim$35 times 
higher than for the reference sample\footnote{The fraction of low-redshift quasars is 
higher for blue selection because the DPOSS faint cutoff is brighter in the red
band (see Table 3).}. The corresponding fraction for region VII (which includes
high-redshift quasars and, possibly, variable stars) is even higher ($\sim$150), 
but the statistics are less robust due to a smaller number of sources. Another quantitative 
representation of the color differences introduced by the variability requirement is shown for 
$u-g$ color in Fig.~\ref{ugHistAll}. These, and other differences listed in Table 
4, demonstrate that the sample of selected candidate variables is {\bf not} dominated 
by spurious objects.

\begin{figure} 
\plotfiddle{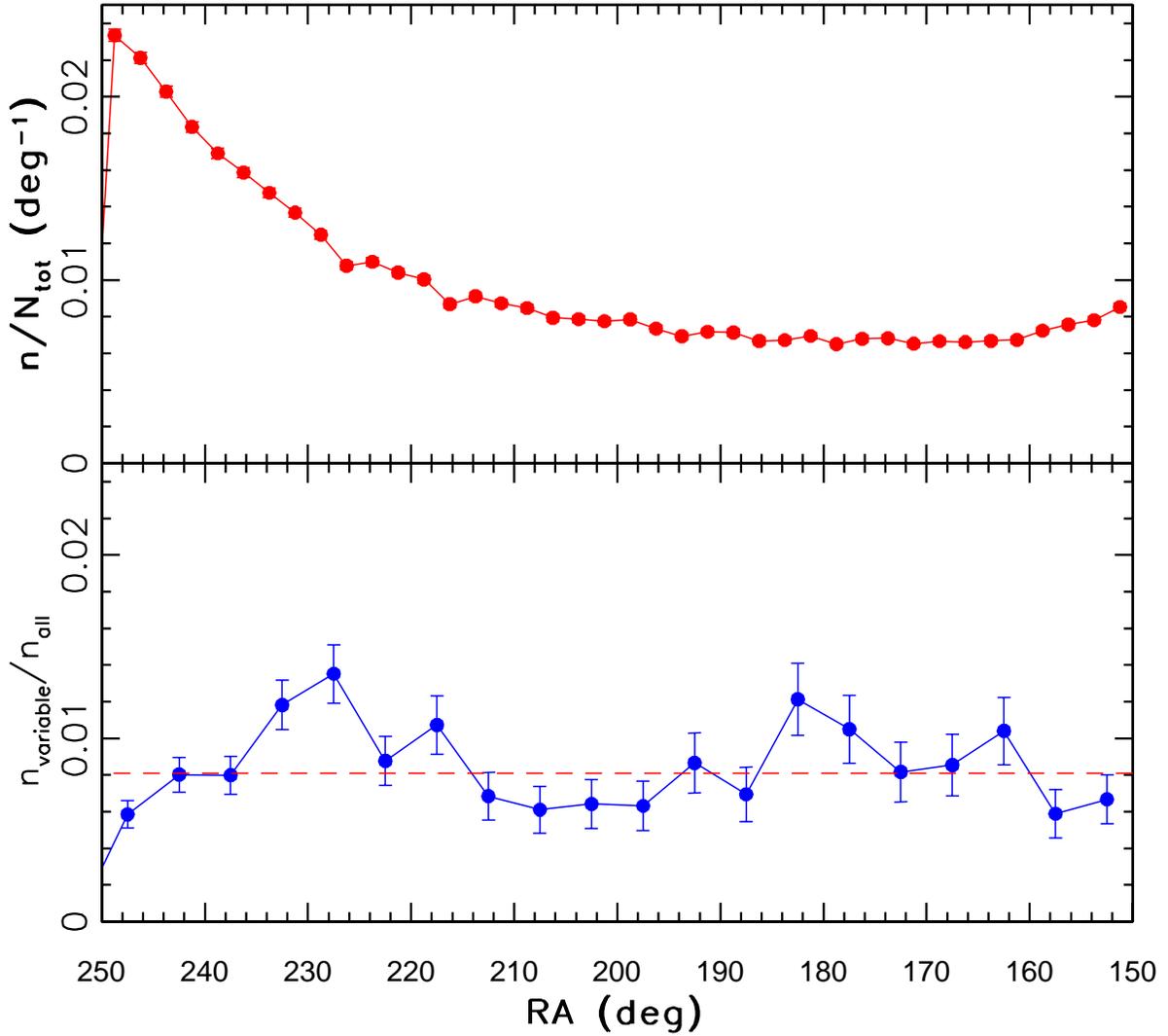}{14cm}{0}{85}{85}{-260}{-220}
\caption{
The counts of all sources (top) and fraction of candidate variables 
in a 2.5 deg. wide strip centered on the Celestial Equator, as a 
function of RA. The counts increase by a factor of $\sim$3 towards the 
left edge because of the decreasing galactic latitude. The fraction 
of candidate variables stays constant (at 0.8\%) within Poissonian 
noise.
\label{RArNGC}
}
\end{figure}

The fraction of selected candidate variables across the sky is stable, and, in particular, 
does not depend on the stellar number density, nor shows jumps at the boundaries of 
Schmidt plates. Figure \ref{RArNGC} illustrates this stability for a 2.5 deg. wide 
strip centered on the Celestial Equator, where the fraction of candidate variables 
remains $\sim$0.8\% (not corrected for unknown selection incompleteness), despite 
the stellar counts varying by a factor of $\sim$3 (a slight increase at RA$\sim$230 
is caused by RR Lyrae in the Sgr dwarf tidal stream, see Section~\ref{RRLyr}).

\subsubsection{ The Comparison with Repeated SDSS Imaging Observations}

\label{SDSSmulti}

The analysis presented in the previous section shows that the selected candidate
variables are not dominated by spurious sources. Here we obtain a quantitative
estimate of the selection efficiency using repeated SDSS imaging data. For about 
10\% of the sky area analyzed here (the SDSS southern equatorial strip, see 
York et al. 2000), there exist between 6 and 9 epochs of SDSS imaging, obtained 
over a period of four years. Both due to a larger number of epochs and to more accurate 
photometry ($\sim$0.02 mag, for details see Ivezi\'{c} et al. 2003b), these data 
have a much higher completeness and efficiency for discovering variable sources 
than the photographic/SDSS presented here. We select variable sources from repeated SDSS 
scans by requiring an rms variability larger than 0.03 mag in the $g$ band. This selection 
results in a negligible fraction of spurious candidates ($<$1\%), a high 
completeness (for example, more than 90\% for RR Lyrae stars), and is also sensitive to 
long-period variables and quasar variability. About 7\% of point sources brighter 
than $g=19$ pass the adopted selection cut (for details see Ivezi\'{c} et al. 2004,
in prep).

\begin{figure}
\plotfiddle{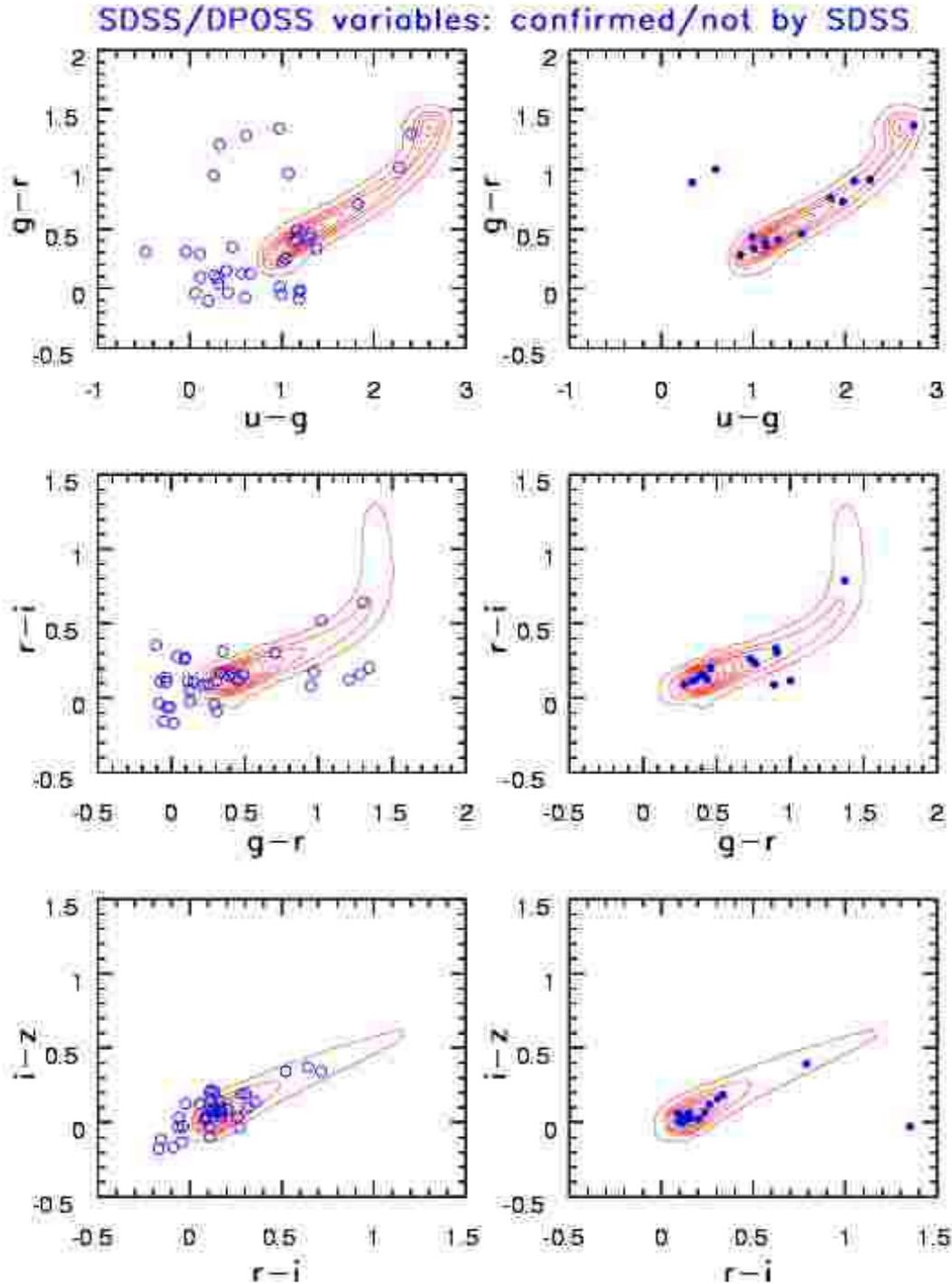}{18cm}{0}{78}{78}{-230}{-50}
\caption{
The distribution of SDSS-DPOSS candidate variable sources (restricted $G$ selection,
see Table 3) confirmed to vary by multi-epoch SDSS imaging (left column, 73\% of the
sample), and those that did not show any evidence for variability (right column, 27\% of
the sample). Note that the latter are mostly found in the stellar locus. 
\label{sdssMulti}
}
\end{figure}

Using the SDSS-DPOSS G-band candidates selected by relaxed criteria (see entry 5 
in Table 3), we find that repeated SDSS scans (at least 6 epochs) exist for 955 sources.
About half of these (49\%) are confirmed as variable by SDSS data. The SDSS-POSS
candidate variables not confirmed as variables by multi-epoch SDSS data tend
to be at the faint end and have smaller SDSS-POSS magnitude differences than
the confirmed variables. Thus, the sample efficiency can be further increased by 
a more restrictive selection requiring $g<19$ and other cuts listed as entry 7
in Table 3. This selection results in 51 SDSS-DPOSS candidate variables with
multi-epoch SDSS data; 73\% are confirmed as variable. The distributions of confirmed
and spurious SDSS-DPOSS variables in SDSS color-color diagrams are compared in 
Fig.~\ref{sdssMulti}. It is not surprising that most of the spurious SDSS-DPOSS variables
are found in the stellar locus, because for a given contamination fraction 
(which is not expected to be a strong function of color) most of the contaminants
come from the most populated part of the diagram. Repeating this analysis 
separately for sources from inside and outside the stellar locus, we find 
that the fraction of true variable sources among the selected candidates from 
the locus is 54\%, while outside the locus it is as high as 94\%. Assuming that 
no more than 10\% of sources from the locus are truly variable, the former fraction 
implies that the decision to tag a source as a candidate variable is correct in more than
95\% of cases (for $g<19$).

\subsubsection{ The Large-amplitude Variables and Visual Comparison of Images}

\begin{figure}
\plotfiddle{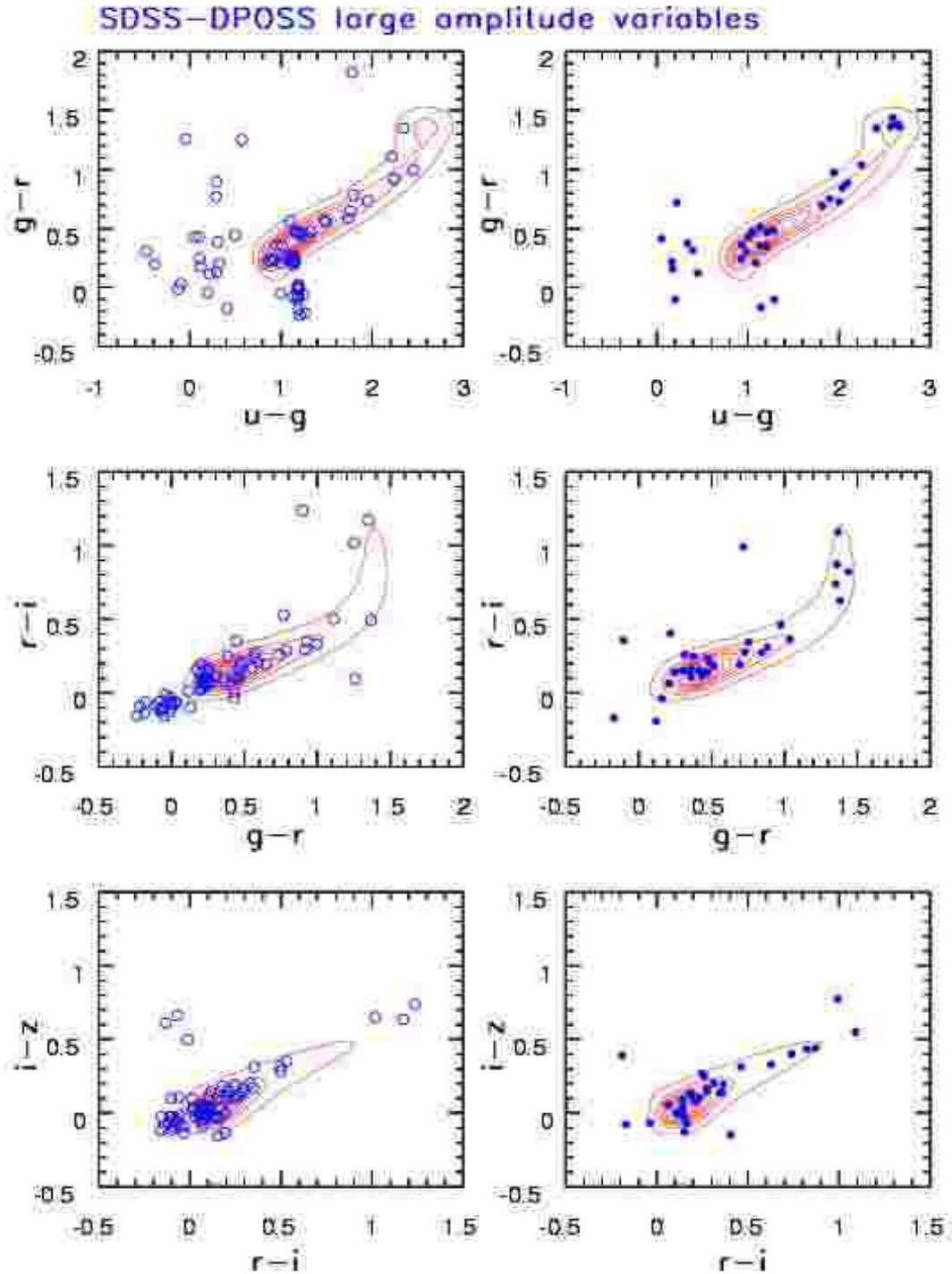}{18cm}{0}{78}{78}{-230}{-50}
\caption{
The distribution of SDSS-DPOSS candidate variable sources with 14.5$<g<$18.5 
and large amplitudes in representative SDSS color-color diagrams.
The symbols show 70 objects with $0.75 < |\Delta G| < 1$ in the left column, 
and 24 objects with $1 < |\Delta G| < 3$ in the right column. The overall 
distributions of SDSS sources with the same magnitude limit are shown by contours. 
\label{selFrac}
}
\end{figure}

\begin{figure}
\plotfiddle{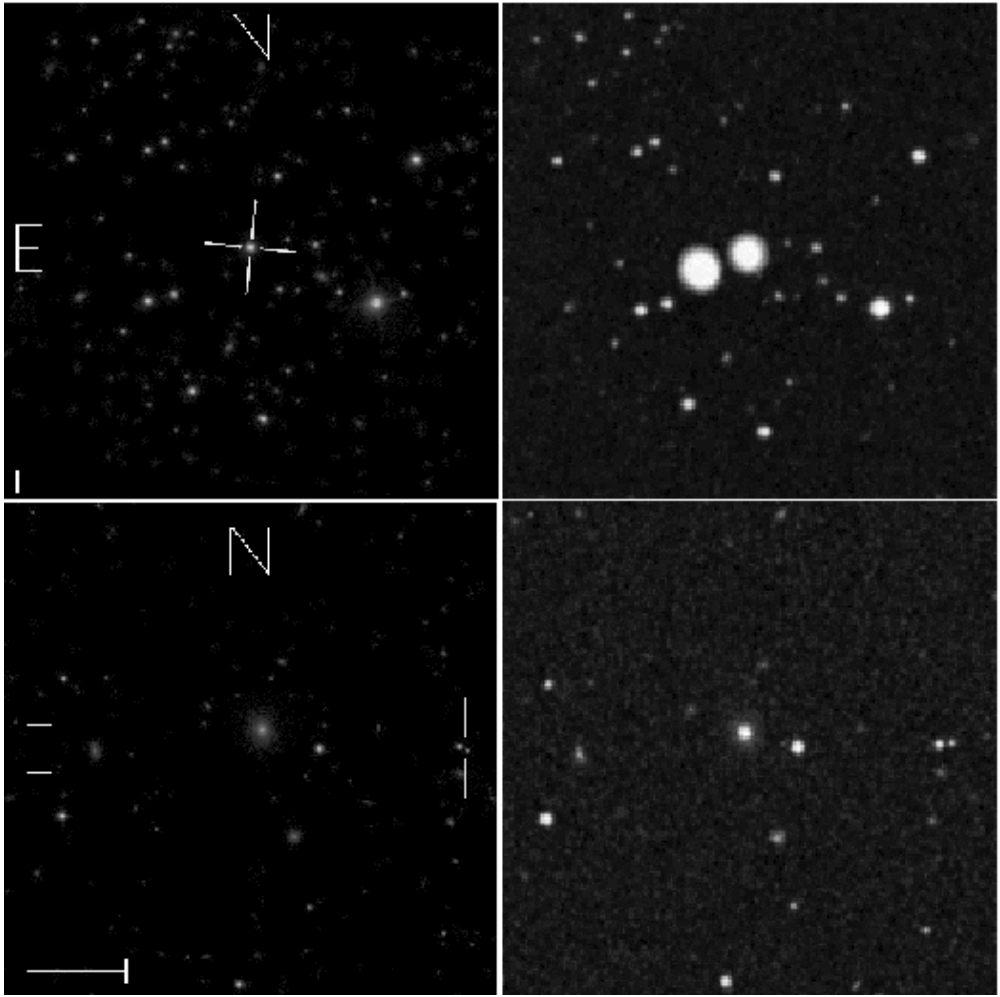}{13cm}{0}{80}{80}{-240}{-100}
\caption{
Examples of spurious candidate variable sources. The left column displays the 
5x5 arcmin $g$ band SDSS images, and the right column displays the blue POSS I 
images on the same scale, and with the same orientation. In the top panels, 
the source marked by a cross was selected as a large amplitude candidate variable. 
The visual inspection of POSS image confirmed that a much brighter source existed 
in the POSS image, as well as another nearby bright source, both of which
turned out to be artefacts. The bottom panels show an example where 
the POSS photometry was noticeably affected by a nearby source (which happened 
to be a fast proper motion object). 
\label{examplesBad}
}
\end{figure}

The presumed large amplitude variables ($\ga$1 mag) may be more likely to be 
spurious (e.g. due to various defects on photographic plates). This possibility 
cannot be robustly tested using methods from the previous section due to insufficient
number of sources. On the other hand, presumed variations with such a large 
amplitude can be tested by the visual comparison of SDSS and POSS images.
The distribution of 70 SDSS-DPOSS candidates with $0.7 < \Delta G < 1$ and 24 
candidates with $1 < \Delta G < 3$ in SDSS color-color diagrams is shown in 
Fig.~\ref{selFrac}. As evident, their distribution does not follow the distribution 
for the reference sample, indicating that they are not dominated by spurious 
candidates. We have visually inspected POSS and SDSS images for these 94 
candidates and found that only $\sim$30 may have been affected by nearby bright
stars. Additional visual inspection of large amplitude variables selected using the 
USNO-A2.0 catalog recovered a spectacular case shown in the top panel in 
Fig.~\ref{examplesBad}. After analyzing the plate print, as well as the 
brightness profiles, we concluded that the two bright POSS sources were
an artefact\footnote{This is a good example of benefits afforded by large
collaborations: they provide an increased statistical chance of working with 
a sufficiently senior member familiar with old technologies and all their pitfalls.} 
(probably caused by splattered liquid on the POSS plate). While this is
a disappointing outcome, it, nevertheless, vividly demonstrates the ability 
of the selection method to recognize differences between POSS and SDSS data. 
Another example of a spurious candidate is shown in the bottom panel in 
Fig.~\ref{examplesBad}. Due to a nearby star, which happened to be a large
proper motion object, the candidate's POSS photometry was noticeably affected, 
while the more accurate SDSS photometry reported a single object with a correct
magnitude. 

\begin{figure}
\plotfiddle{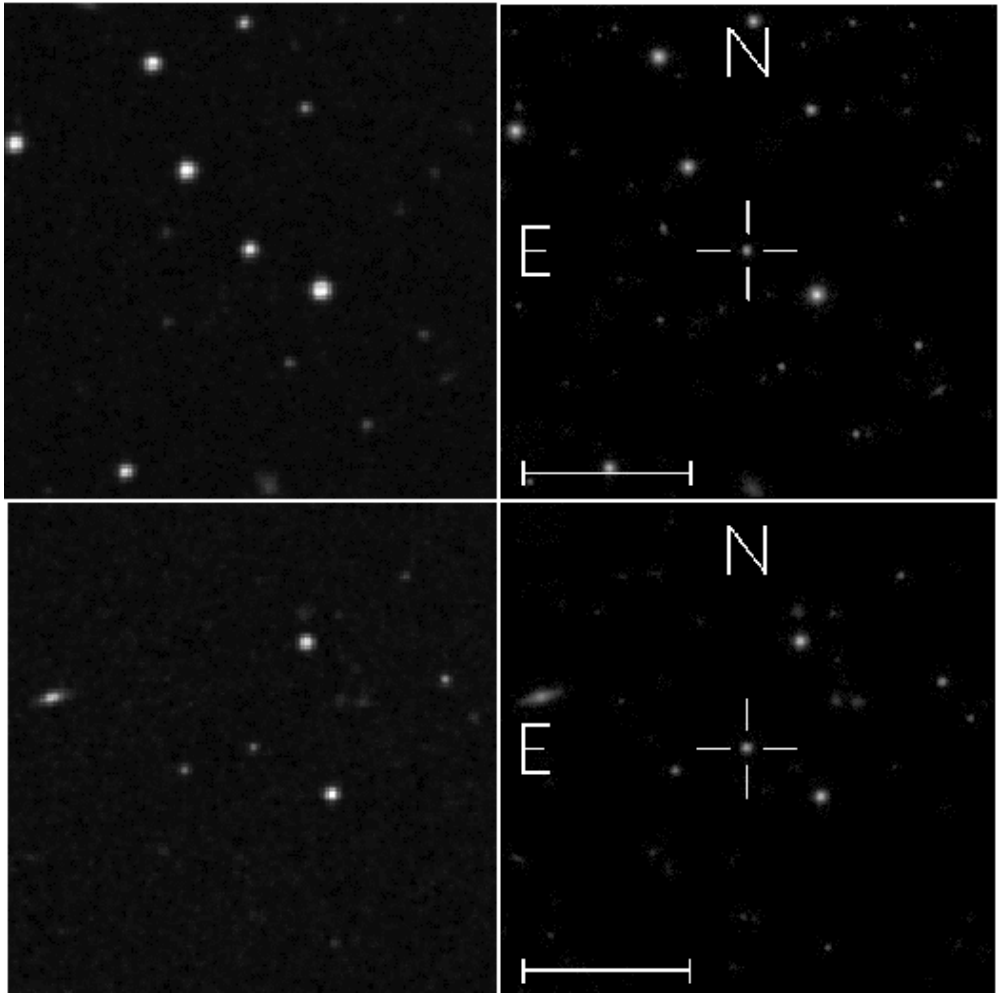}{13cm}{0}{80}{80}{-240}{-100}
\caption{
Examples of large amplitude ($\sim1.5$ mag) candidate variable sources. The right 
column displays the 3x3 arcmin $g$ band SDSS images, and the left column displays 
the blue POSS II images on the same scale, and with the same orientation. The sources
marked by a cross are clearly variable. The top source was brighter in POSS, 
and the bottom source in SDSS. 
\label{examplesGood}
}
\end{figure}

Despite these pitfalls, the SDSS-POSS comparison does yield true large amplitude
variables. For example, one of the sources with an SDSS-POSS magnitude difference 
of $\sim$2 mag is in the region multiply observed by SDSS (12 epochs). The 
available SDSS data demonstrate that it is a long-period variable with a 
peak-to-peak amplitude exceeding 5 mag (Ivezi\'{c} et al. 2004, in prep). 
This star, and another example of a large-amplitude variable, are shown in 
Fig.~\ref{examplesGood}.

\section{ The Milky Way Halo Structure Traced by Candidate RR Lyrae Stars     }
\label{RRLyr}

As recently shown (Ivezi\'{c} et al. 2000, Vivas et al. 2001, Ivezi\'{c} et al. 
2003cd), faint RR Lyrae stars have a very clumpy distribution on the sky (most 
prominent features are associated with the Sgr dwarf tidal stream). This 
substructure offers a test of the spatial homogeneity of the selection 
algorithm: the known clumps ought to be recovered to some extent by the 
candidate RR Lyrae stars selected here, if the selection algorithm is robust.  
Furthermore, if such robustness can be demonstrated, the SDSS-POSS candidates
can be utilized to quantify the halo substructure in the areas of sky 
for which multi-epoch SDSS data do not exist.

\subsection{ The $u-g$ Color Distribution of Candidate RR Lyrae Stars}

\begin{figure}
\plotfiddle{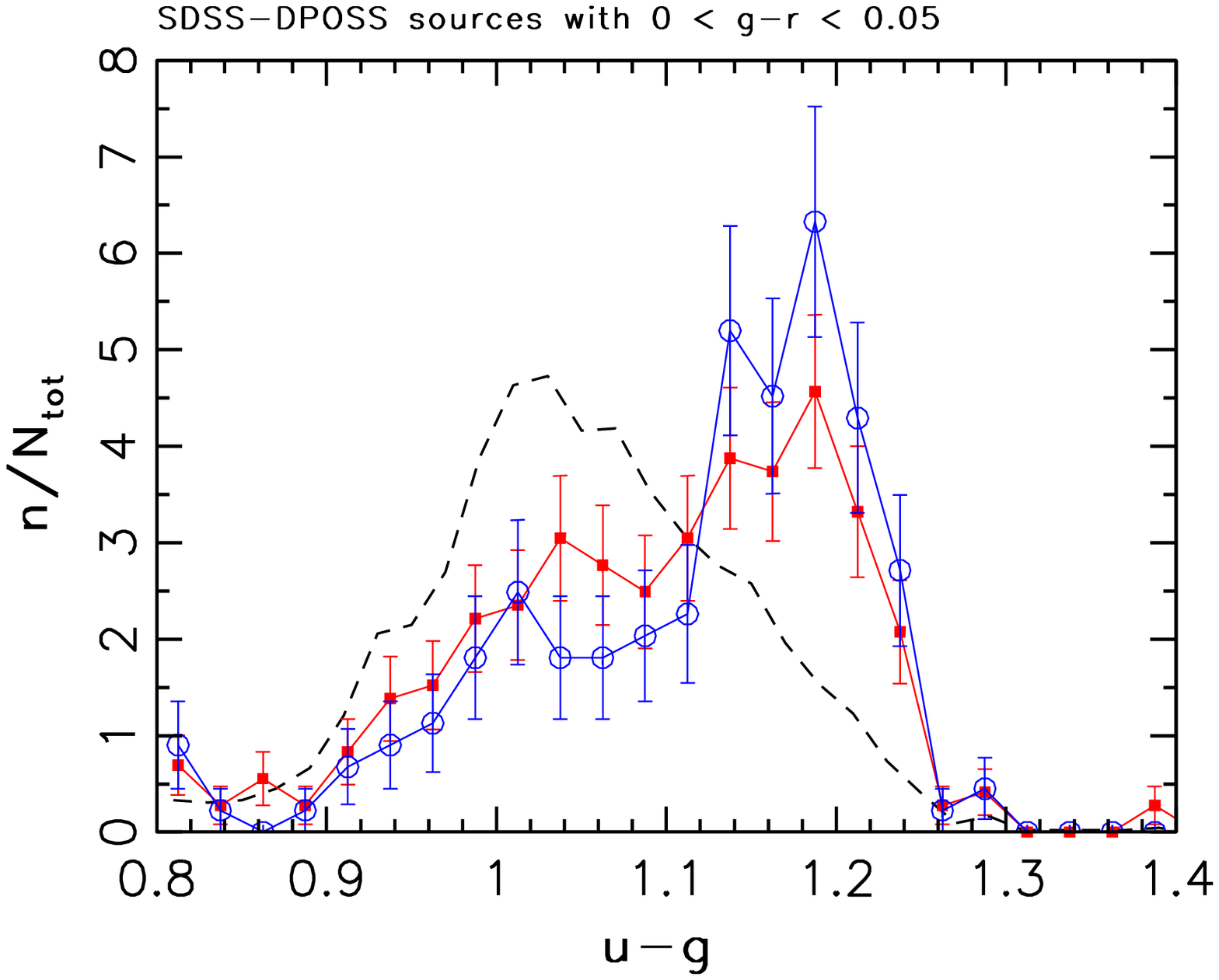}{9cm}{0}{70}{70}{-220}{-240}
\plotfiddle{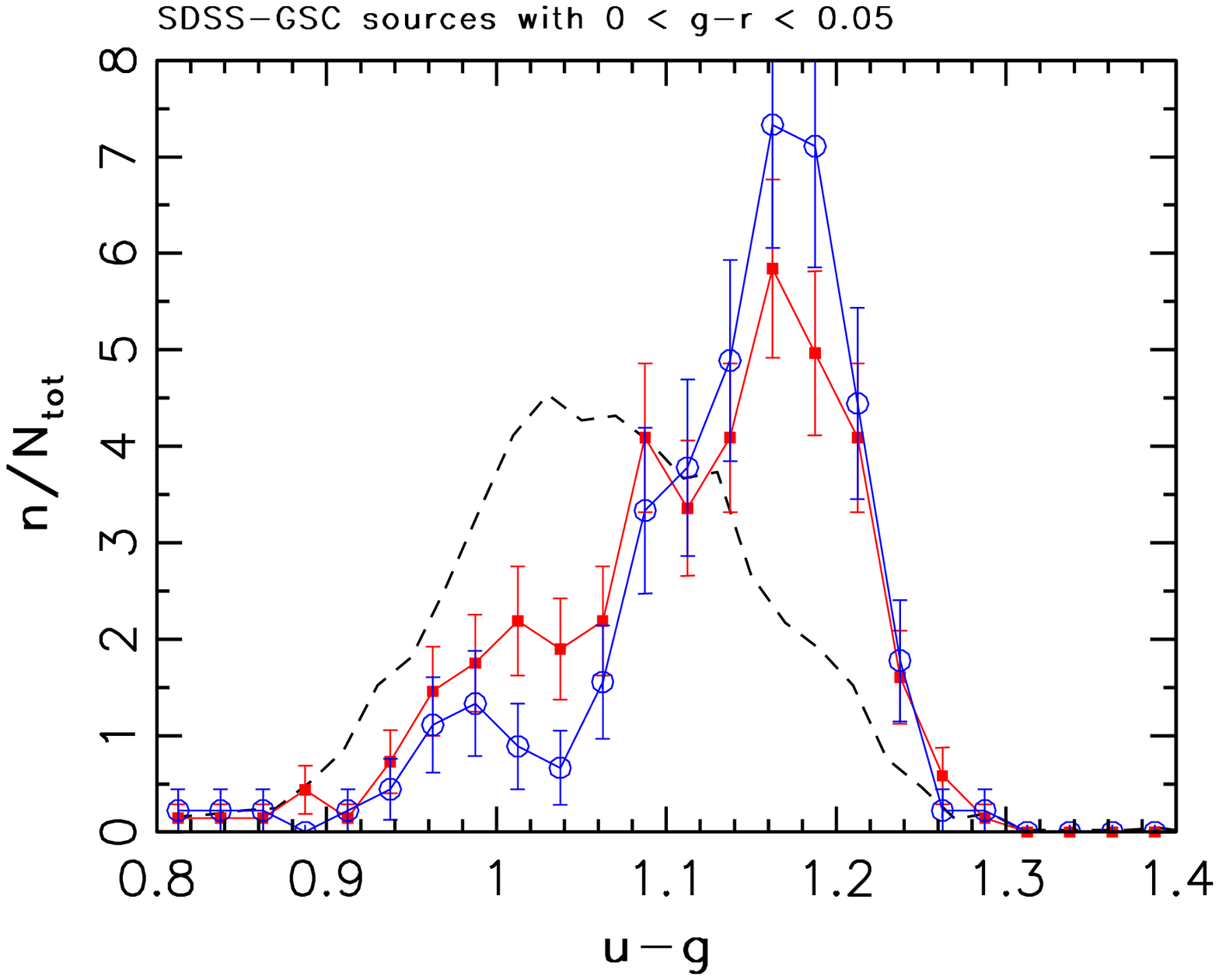}{9cm}{0}{70}{70}{-220}{-260}
\caption{
The comparison of $u-g$ distributions in the range characteristic for RR Lyrae
stars, for candidate variables (symbols with error bars, analogous to Fig.~\ref{ugHistAll}),
and for a reference sample (dashed line), for sources with $0<g-r<0.05$ and $u<20.5$
(top panel for DPOSS-based selection, bottom panel for GSC; circles for selection 
in the blue band, squares for the red band). Note that variable objects, dominated by 
RR Lyrae stars, have redder $u-g$ colors than the reference (full) sample.
\label{grStrip}
}
\end{figure}

Before proceeding with the analysis of spatial distributions of candidate
RR Lyrae stars, we test their selection robustness using a method 
introduced\footnote{This method was suggested to Ivezi\'{c} et al. by 
the referee Abi Saha.} by Ivezi\'{c} et al. (2000). RR Lyrae stars have 
somewhat redder $u-g$ color
($\sim$0.2 mag) than stars with similar effective temperature (i.e. $g-r$ color) 
that are not on the horizontal giant branch. Since the $u$ band flux is not used
in the selection of variable objects (all POSS bands are redder than the SDSS $u$ band), 
this offset is a robust indication that the candidate variables are dominated by true RR
Lyrae stars. Fig.~\ref{grStrip} compares the $u-g$ color distribution for candidate
variable objects to the distribution for all sources in a narrow $g-r$ range 
(0 to 0.05, designed to exclude the main stellar locus). As evident, the selected 
candidates have redder $u-g$ color than the full sample, in agreement with 
the color distribution of RR Lyrae selected using light-curves obtained by 
the QUEST survey (Ivezi\'{c} et al. 2003a). The difference is more 
pronounced for the selection in blue bands (because the variability amplitude 
decreases with wavelength), and somewhat more pronounced for the GSC catalog 
than for the DPOSS catalog. The counts of selected candidates are consistent
with the conclusion from the previous section that the decision to tag a source 
as a candidate variable is correct in more than 95\% of cases.  

We determined the selection completeness for RR Lyrae stars using a 
complete sample of 162 RR Lyrae stars discovered by the QUEST survey 
and discussed by Ivezi\'{c} et al. (2003a). We find that 32 (20\%) of 
these stars are recovered by the relaxed DPOSS G-band selected sample.

\subsection{ The Spatial Distribution of Candidate RR Lyrae Stars}

\begin{figure}
\plotfiddle{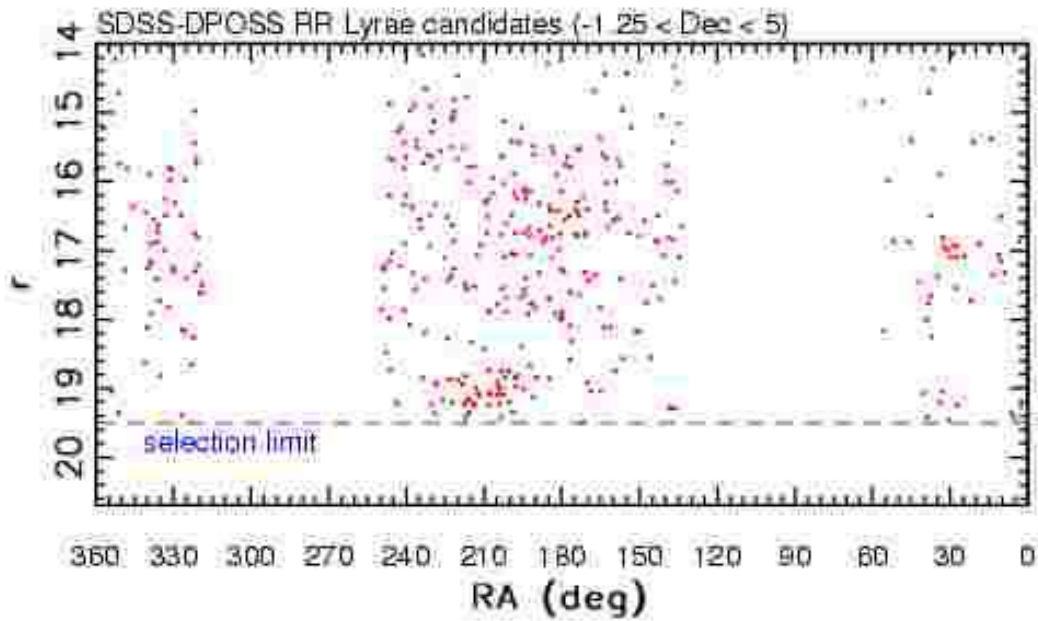}{8cm}{0}{85}{85}{-260}{-400}
\caption{
The magnitude-position distribution of 350 SDSS-DPOSS RR Lyrae candidates within
5 deg. from the Celestial Equator. The sample completeness is fairly uniform for 
$r<19$ and decreases with $r$ towards the selection faint limit of $r=19.5$. 
The clumps easily discernible at (RA,$r$)$\sim$(210,19.2) and at (30,17) are 
associated with the Sgr dwarf tidal stream. The clumps at (180,16.5), and at (330,17) 
have also been previously reported. The clump at ($\sim$235,$\sim$15.5) is a new 
detection.
\label{RArEQdposs}
}
\end{figure}

\begin{figure}
\plotfiddle{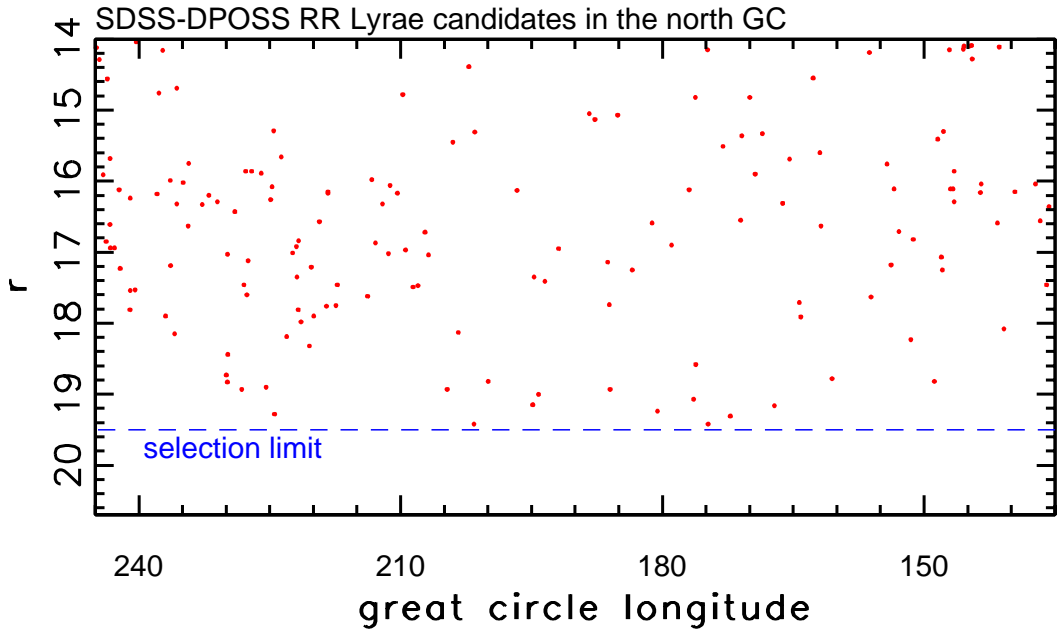}{8cm}{0}{85}{85}{-260}{-400}
\caption{
Same as Fig.~\ref{RArEQdposs}, except that the 161 candidates are selected 
from a 10 deg. wide strip centered on a great circle defined by a 
node at RA=95$^\circ$ and inclination of 65$^\circ$ (for more details about
great circle coordinates see Pier et al. 2003). Note the very inhomogeneous
structure, and, in particular, the fairly prominent feature at the longitudes 
210-240 (RA$\sim$240$^\circ$ and Dec$\sim$50$^\circ$), with $r \sim$16-18.
\label{RArNGCdposs}
}
\end{figure}

Using the relaxed DPOSS G-band selected sample (GSC-based selection performs 
better, but the public catalog is not deep enough to probe the outer halo), 
we isolate 628 RR Lyrae candidates by adopting color boundaries from Ivezi\'{c} 
et al. (2003a). The magnitude-position distribution of 350 candidates within 5 deg. 
from the Celestial Equator is shown in Fig.~\ref{RArEQdposs}.
The sample completeness is fairly uniform for $r<19$ and decreases with $r$ towards 
the selection faint limit of $r=19.5$ (corresponding to $\sim$60 kpc). The clumps 
easily discernible at ($\alpha_{2000}$,$r$)$\sim$(210,19.2) and at (30,17) are associated with 
the Sgr dwarf tidal stream. The clumps at (180,17) and at (330,17) have also been 
previously reported (Vivas et al. 2001, Ivezi\'c et al. 2003c). The recovery of 
these known structures suggests that the clump at ($\sim$235,$\sim$15.5), which has 
not been previously reported, is a robust detection. Another previously unrecognized 
clump is detected around $\alpha_{2000}$$\sim$240$^\circ$ and $\delta_{2000}\sim$50$^\circ$ 
(see Fig.~\ref{RArNGCdposs}). The significance of these newly recognized structures 
will be placed in a broader context of other available data elsewhere (Ivezi\'{c} et al.,
2004, in prep.).

\section{            The Long-term Variability of Quasars              } 
\label{varqso}

The optical continuum variability of quasars has been recognized since 
their first optical identification (Matthews \& Sandage 1963), and it has 
been proposed and utilized as an efficient method for their discovery 
(van den Bergh, Herbst, Pritchet 1973; Hawkins 1983; Hawkins \& Veron 1995; 
Ivezi\'{c} et al. 2003e). The observed characteristics of the variability of
quasars are frequently used to constrain the origin of their emission 
(e.g. Kawaguchi et al. 1998, and references therein; 
Martini \& Schneider 2003). 

Recently, significant progress in the description of quasar variability has 
been made by employing SDSS data (de Vries, Becker \& White 2003, hereafter dVBW;
Vanden Berk et al. 2004, hereafter VB). The size and quality 
of the sample analyzed by VB (two-epoch photometry for 25,000 
spectroscopically confirmed quasars) allowed them to constrain 
how quasar variability in the rest frame optical/UV regime depends upon 
rest frame time lag, luminosity, rest wavelength, redshift, the presence 
of radio and X-ray emission, and the presence of broad absorption line outflows.
However, the time lags probed by the available SDSS data (up to 3 years) 
are too short to detect deviations of the structure function 
(the root-mean-square scatter of measured magnitudes, see Eq. 1 in dVBW) from 
a simple power-law that are expected for long time lags (Cid Fernandes, 
Sodr\'{e}, \& Vieira da Silva 2000, and references therein).

The much longer time lags between POSS and SDSS ($\sim$50 years in the observer's
frame) offer the possibility of detecting such deviations, despite larger photometric
errors for the POSS catalogs, and to study the long-term characteristics of 
quasar variability. Using a recalibration approach similar to the
one described here (except that only fields around known quasars were 
recalibrated), dVBW studied long-term variability for 3,791 quasars from 
the SDSS Early Data Release (Stoughton et al. 2002). Here we have assembled a larger 
sample of SDSS quasars ($\sim$17,000, Schneider et al. 2003), which allows 
us to constrain the overall shape of the $m_{SDSS}-m_{POSS}$ distribution 
and not only its root-mean-square scatter, as discussed in the next section. 
The dependence of the long-term quasar variability on luminosity, rest-frame 
wavelength, and time lag is analyzed in the subsequent section.

\subsection{ The Distribution of SDSS-POSS Magnitude Differences for Quasars }
 
Analysis of the multi-epoch SDSS imaging data suggests that the distribution of 
$\Delta m$ for quasars is better described by an exponential distribution than by 
a Gaussian distribution, for all bands ($ugriz$) and time scales probed (up to 
4 years time lag in the observer's frame). This result is independent of whether the data are 
binned by wavelength and time lag in the rest or observer's frame (Ivezi\'{c} et al. 
2004, hereafter I04, in prep). Here we investigate whether this result can be 
reproduced for much longer time lags using SDSS-POSS measurements.

\begin{figure}
\plotfiddle{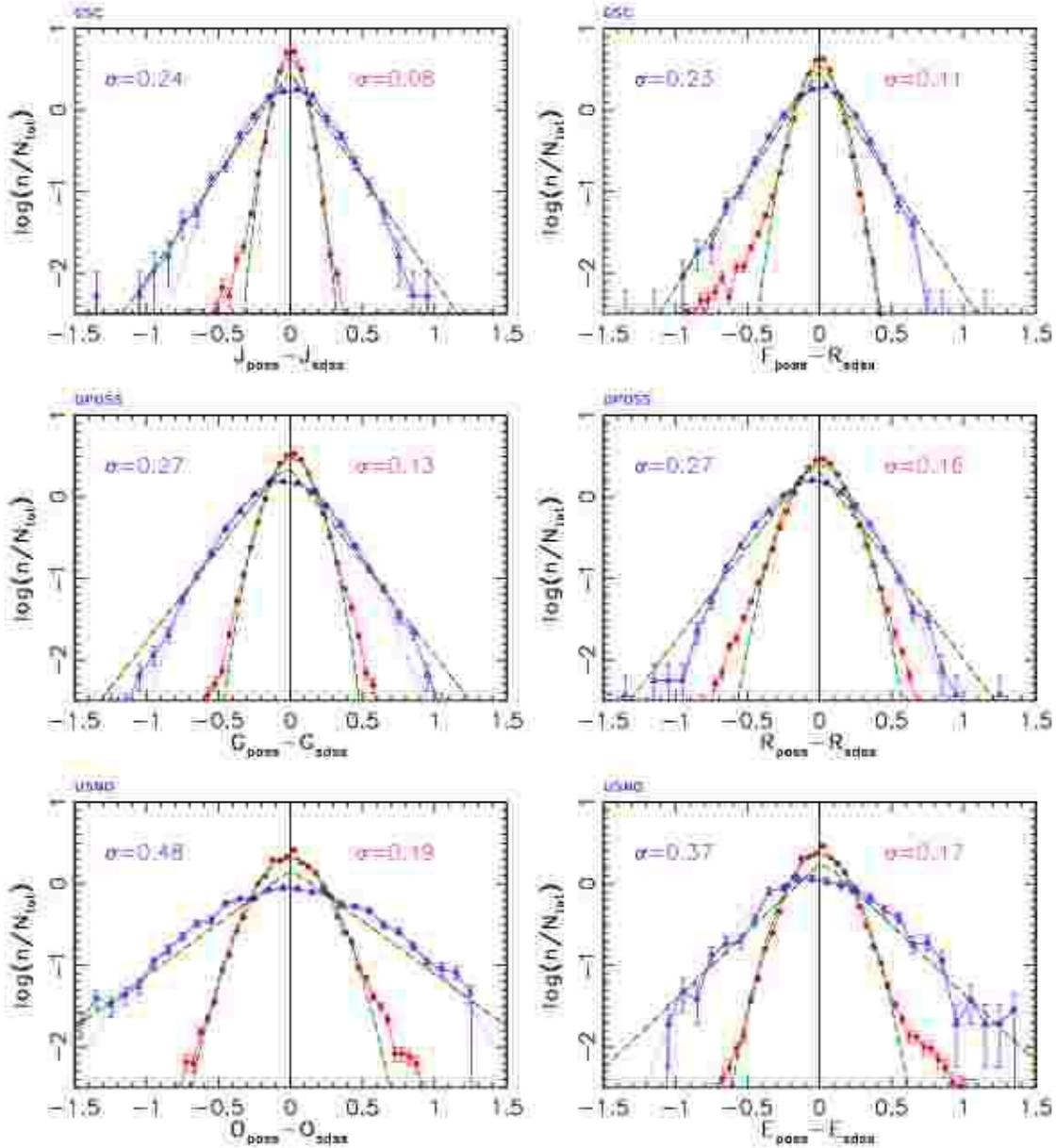}{18cm}{0}{80}{80}{-260}{-160}
\caption{
The probability density distributions of POSS-SDSS magnitude differences for 
color-selected low-redshift quasars is shown by triangles, and compared to the 
corresponding distribution for stars, shown by large dots (note
logarithmic scale). The rms for each distribution measured using 
interquartile range is also shown in each panel. The solid lines show Gaussian 
distributions that have the same rms scatter as the data. Note that
magnitude differences for stars are well described by a Gaussian. The 
dot-dashed lines show an exponential distribution that has the same rms 
as the data for quasars, and seem to provide a marginally better fit
than a Gaussian distribution. 
\label{dmQSO}
}
\end{figure}

Fig.~\ref{dmQSO} shows the magnitude difference distributions for stars and 
spectroscopically confirmed quasars with redshifts in the range $0.3<z<2.4$, measured 
using GSC, DPOSS, and USNO-A2.0 catalogs. The distributions for quasars are marked 
by triangles, and those for a control sample of stars with the same magnitude 
distribution\footnote{The slope of the differential magnitude distribution (``number counts'') 
for quasars is much steeper than that for stars. Since the photometric errors 
increase with magnitude, care must be taken to properly account for the error
contribution to the measurement of the structure function. A simple comparison
of two magnitude-limited samples of stars and quasars results in an underestimated
photometric error contribution to the structure function.} are marked by solid 
circles.

The dashed lines in Fig.~\ref{dmQSO} show exponential distributions that have the 
same rms scatter as the data (the rms for each distribution are shown in the panels, 
and also listed in Table 5), and the dot-dashed lines show Gaussian distributions,
both convolved with a Gaussian of the same width as the distribution of magnitude
differences for stars. While the data presented here do not constrain the tails 
of the $\Delta m$ distributions as well as multi-epoch SDSS data (due to larger photometric 
errors), the obtained magnitude distributions are consistent with the inferences 
made using multi-epoch SDSS data. Typically, $\sim$1\% of the sample is outside
the $\pm 3\sigma$ boundaries, a fraction about a five times larger than expected 
for a Gaussian distribution. While formally significant, it is possible that 
the remaining calibration problems with POSS catalogs have contributed to this
deviation from a perfect Gaussian distribution. In any case, the deviations
are sufficiently small for the rms width to be an efficient statistic for
describing the observed distributions.

These observed rms values, listed in Table 5, are $\sim$1$\sigma$ ($\sim$0.05-0.10 mag) 
smaller than the values obtained by dVBW. The photometric errors (i.e. structure function 
for stars) displayed in Fig. 8 from dVBW correspond to the smaller of the two curves 
shown in their Fig. 4. Adopting the other curve decreases the estimate of the quasar 
variability as measured by dVBW, and thus decreases the discrepancy with our results 
to a $<$1$\sigma$ level. It is noteworthy that the magnitude difference distributions 
for stars shown in Fig.~\ref{dmQSO} have smaller rms values than do the 
structure functions for stars shown in Fig. 4 from dVBW (our values correspond
to log(SF)$\sim$-0.65 or less). Thus, it is plausible that the remaining slight 
discrepancy is due to somewhat different procedures used to recalibrate POSS catalogs.

\subsection{ The Turn-over in the Structure Function }

The extrapolation of the power-law dependence of the quasar rms variability on time
measured at short time scales using repeated SDSS imaging (I04) predicts that the 
quasar rms variability measured using SDSS and POSS I should be of the order 0.60 mag, 
and 0.35 mag for SDSS-POSS II. Since the measured values (see Table 5) are smaller 
than these extrapolated values, they present strong evidence for a turn-over in the 
quasar structure function. 

\begin{figure}
\plotfiddle{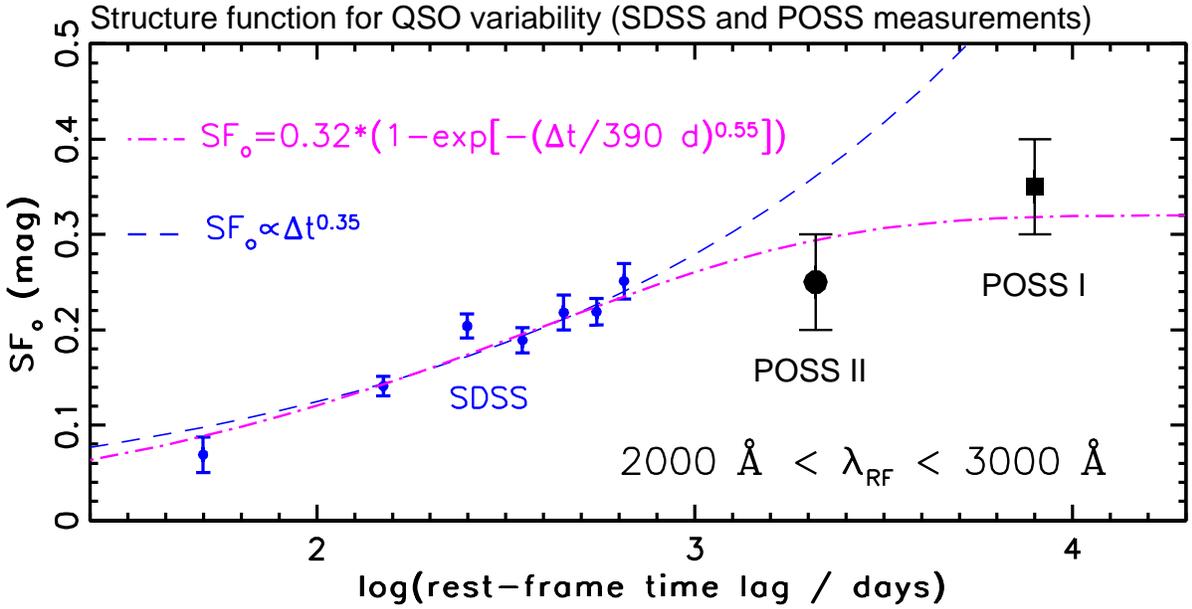}{8cm}{0}{90}{90}{-288}{-430}
\caption{The long-term dependence of structure function on 
rest-frame time lag, in the range 2000--3000 \AA, for two data 
sets: SDSS-SDSS for short time lags (small symbols, adopted
from I04), and SDSS-POSS for long time lags (large symbols). The 
observed SDSS-POSS long-term 
variability is smaller than predicted by the extrapolation of 
the power-law measured for short time scales using repeated SDSS 
imaging (dashed line): the measured values are 0.35 and 0.24 mag, 
while the extrapolated values are 0.60 and 0.35 mag for SDSS-POSS I 
and SDSS-POSS II, respectively. The dot-dashed line shows a
simultaneous best-fit to all the displayed data.
\label{SFQSO}
}
\end{figure}

Fig.~\ref{SFQSO} shows the dependence of structure function on rest-frame time lag, 
in the range 2000--3000 \AA, for two data sets: SDSS-SDSS for short time lags (small 
symbols, I04; note that the variability inferred from 
repeated imaging scans is fully consistent with the results presented by VB, that were
based on a comparison of imaging and spectrophotometric magnitudes), and SDSS-POSS 
for long time lags (large symbols). The extrapolation of the power law measured for 
short time scales (dashed line) clearly overestimates the amplitude of the structure
function reported here. We fit the observed dependence of structure function on
rest-frame time lag using the following functional form
\begin{equation}
  SF(\Delta t_{RF}) = D\,
  \left(1 - {\rm e}^{-(\Delta t_{RF} / \tau)^\gamma}\right)    
\end{equation}

The best fit parameters are $D=0.32\pm0.03$, $\tau=(390\pm80)$ days,
and $\gamma=0.55\pm0.05$. This best-fit is shown in Fig.~\ref{SFQSO}
by the dot-dashed line. We conclude that {\it the characteristic time 
scale for optical variability of quasars is of the order 1 year in the
rest frame.} We postpone a more detailed analysis to a forthcoming paper.

\subsection{ The Dependence of Long-term Variability on Luminosity, 
                        Rest-frame Wavelength and Time Lag }

The distribution of time differences in the observer's frame for SDSS-POSS variability 
measurements is strongly bimodal ($\sim$10 years for SDSS-POSS II and $\sim$50 years for
SDSS-POSS I comparison). However, the distribution of time lags measured in the quasar 
rest-frame is more uniform due to a wide distribution of redshifts (the $1+z$ effect; 
for a discussion, see dVBW). The same effect also widens and flattens the distribution 
of rest-frame wavelengths. For the data discussed here, the rest-frame time lags span 
the range 1,200-13,500 days, and the rest-frame wavelength is limited to the range 
1,400-4,800 \AA\ (using the bluest two bands in each survey).  

In this section we investigate whether the correlations of variability with
luminosity and rest-frame wavelength observed for short time lags (VB, I04) 
are valid for long time lags. In particular, we examine whether the decrease of
variability with luminosity and wavelength, and increase with time, are observed
for large time lags. 

The dependence of variability on numerous relevant parameters complicates the analysis, 
and has resulted in many conflicting results in the literature (for a summary 
see Giveon et al. 1999). 
Furthermore, many of these parameters are highly correlated in flux-limited quasar 
samples due to steep quasar ``counts'' (the differential apparent magnitude distribution). 
Nevertheless, the sample considered here is sufficiently large that some of these 
degeneracies can be lifted by simple binning, similarly to the analysis described by VB.

\begin{figure}
\plotfiddle{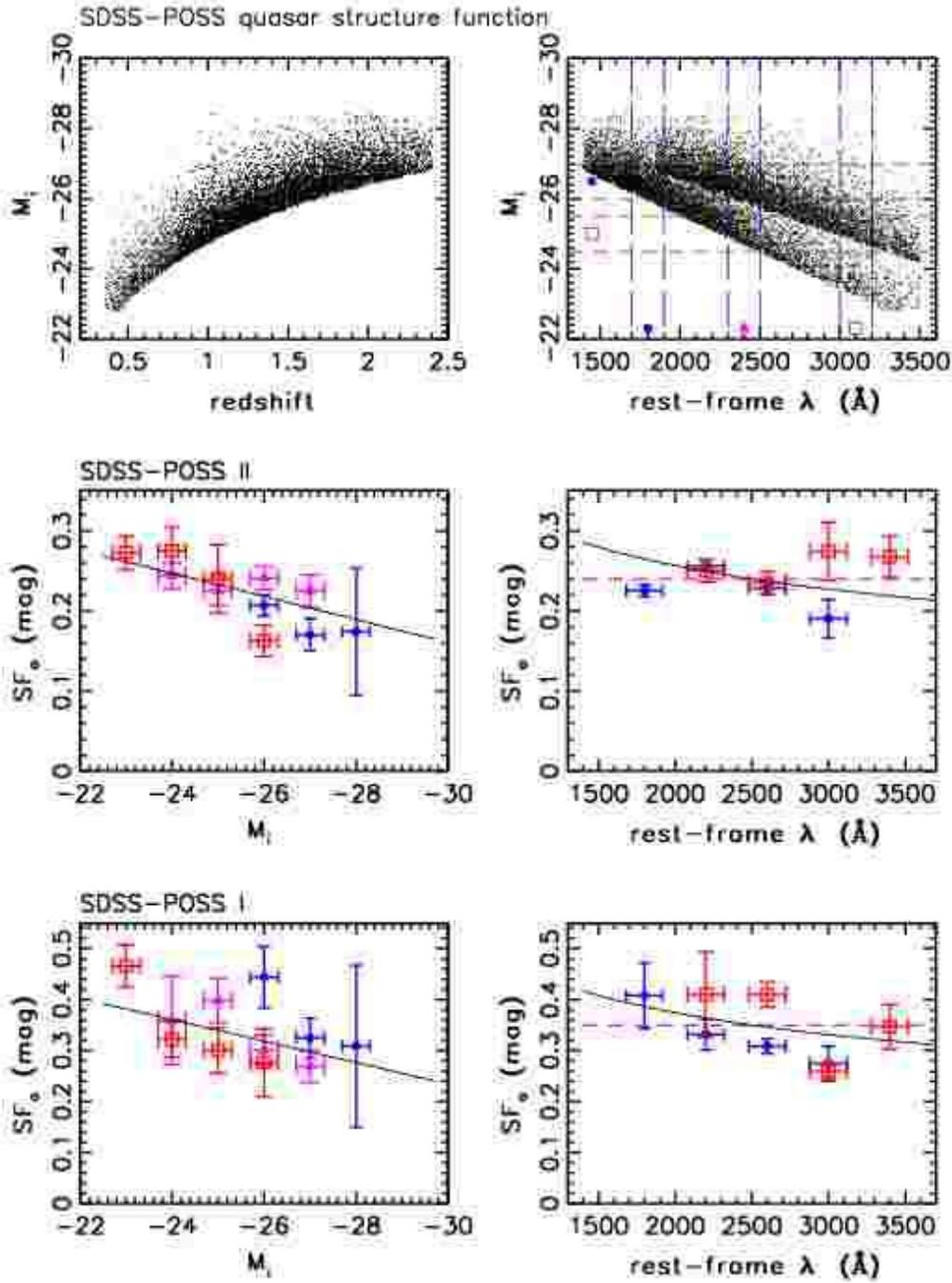}{17cm}{0}{76}{76}{-250}{-60}
\caption{The top two panels show the distribution of 7,279 quasars with 
$i<19$, $G_{SDSS}<19$, and redshifts in the range 0.3--2.4, in the redshift 
vs. $M_i$ (left) and $M_i$ vs. rest-frame wavelength (right) planes. 
The remaining four panels show structure function as a function of $M_i$ 
and rest-frame wavelength for objects selected in narrow bins marked in 
the top right panel (the symbols in each strip are used to mark the 
corresponding histograms). The middle two panels correspond to SDSS-POSS II 
comparison, and the bottom two panels to SDSS-POSS I comparison. 
The solid lines in the two left panels show the correlation $SF \propto 1+0.024\,M_i$, 
inferred by I04 from repeated SDSS imaging scans. The dashed lines
in the two right panels represent median values, and the solid lines
are the relationship $SF\propto \lambda^{-0.3}$, derived from repeated 
SDSS imaging data. See text for discussion.  
\label{SFpossRF}}
\end{figure}

The top left panel in Fig.~\ref{SFpossRF} shows the distribution of 7,279 quasars with 
$i<19$, $G_{SDSS}<19$, and redshifts in the range $0.3<z<2.4$, in the redshift vs.
$i$ band absolute magnitude plane ($M_i$, computed using the WMAP cosmological 
parameters, $\Omega_m=0.27$, $\Omega_\Lambda=0.73$, $h=0.71$, and K-corrected using 
$F_\nu\propto \nu^{-0.5}$). Their distribution in the $M_i$ vs. rest-frame wavelength 
plane is shown in the top right panel (the two overlapping distributions arise from
the two POSS bands used in this analysis, $O$ and $E$ for POSS I, and $G$ and $R$ for
POSS II). 

We decouple the variability dependence on absolute magnitude and rest-frame 
wavelength by computing the structure function for objects selected in narrow
bins, shown in the top right panel in Fig.~\ref{SFpossRF} by short-dashed lines 
(two bins with nearly constant $M_i$) and long-dashed
lines (three bins with nearly constant $\lambda$). The symbols in each strip are
used to mark the corresponding histograms in the lower four panels. The dependence
of the structure function on rest-frame time lag is essentially the same as
its dependence on rest-frame wavelength because the ratio of these two 
quantities is nearly constant ($\Delta t_{RF} \sim C \lambda_{RF}$), with 
$C\sim0.8$ days/\AA\ for SDSS-POSS II data, and $C\sim3.2$ days/\AA\ for 
SDSS-POSS I data (that is, $\lambda_{RF}$ in the two lower right panels in 
Fig.~\ref{SFpossRF} is nearly equal to $\Delta t_{RF} / C$).

The middle left panel shows the dependence of the structure function (computed using
the definition from dVBW, and corrected for measurement errors) on $M_i$
for SDSS-POSS II magnitude differences. The DPOSS bands (we used $G$ and $R$ bands
in this analysis) produce consistent results, which agree well with the correlation 
$SF \propto 1+0.024\,M_i$ (shown by the line), inferred by I04 from repeated SDSS 
imaging scans.\footnote{This expression agrees well with the result from VB who used 
a different functional form to describe this correlation.}
The middle right panel shows the dependence of the structure function on
the rest-frame wavelength.  The data are fully consistent with a constant value, 
shown by the dashed line (0.24 mag, median of all the points). This behavior is 
in contrast with the relationship $SF\propto \lambda^{-0.3}$, derived from repeated 
SDSS imaging data (I04). However, this result is affected by the correlation 
in the SDSS-POSS II sample between rest-frame wavelength and time lag --- 
the increase of variability with time for the rest-frame time lags probed 
($\sim$1300-3400 days) offsets its decrease with rest-frame wavelength. 

This effect is expected to be much weaker in SDSS-POSS I data due to a 
turn-over in the quasar structure function discussed in the previous section.
Indeed, as the bottom two panels demonstrate, while the dependence of variability 
on $M_i$ is still reproduced (albeit with larger errors), the variation of 
structure function with wavelength is consistent with $SF\propto \lambda^{-0.3}$. 
Note that we did not correct the data points for the systematic residual dependence 
on rest-frame wavelength in the two bottom left panels (the correction is 
$\pm0.03$ mag for the two edge bins, relative to the middle one), and on absolute 
magnitude in the two bottom right panels (relative offset between the two 
bins is 0.02 mag). We conclude that there is no evidence that the dependence of 
the structure function on luminosity and rest-frame wavelength is different for 
long time lags discussed here (4-40 years in the rest frame) than for shorter 
time lags (VB, I04). A more detailed analysis of these data is postponed to a 
future publication.

\section{                          Discussion                             } 

We present a direct comparison of photometric measurements available
in public POSS catalogs. The most accurate photometry is provided by the
GSC2.2 catalog. The results of photometric recalibration based on a dense
grid of calibration stars measured by the SDSS demonstrate that errors in POSS 
photometry can be reduced by about a factor of two. POSS I magnitudes can 
be brought to $\sim$0.15 mag accuracy, and POSS II magnitudes to $\sim$0.10 mag
accuracy. While these apparently irreducible errors are considerably larger 
than those delivered by modern CCD data ($\sim0.02$), the POSS catalogs are, 
nevertheless, invaluable for studying sources variable on long time scales. A particular 
success of the recalibration method is that the resulting error distribution 
for POSS photometry is nearly Gaussian, which greatly helps in the design of 
robust algorithms for selecting candidate variable sources.

We designed and tested algorithms for selecting candidate variable sources using 
POSS and SDSS photometric measurements. The algorithm's decision to tag a 
source as a candidate variable is correct in more than 95\% of cases. 
The selection criteria can be tuned up to result in samples that contain no
more than 30\% of spurious candidates. 
  
This is the first study that examined the distribution of sources variable on 
long time scales in SDSS color-color diagrams. Even with the fairly large cutoffs 
for selecting candidate variables (0.20-0.35 mag), we find that at least 1\% of 
faint optical sources appear variable. About 10\% of the variable population are 
quasars, although they represent only 0.25\% of all point sources (for $g<19$).  

Using a sample of $\sim$17,000 spectroscopically confirmed quasars, we demonstrate
that the power-law increase of the quasar variability (structure function, rms) 
with time lag observed for short time lags cannot be extrapolated beyond a few 
years in the rest frame -- such extrapolation predicts a variability level 
significantly larger than measured (0.35 vs. 0.60 mag for SDSS-POSS I, and 
0.24 vs. 0.35 mag for SDSS-POSS II). The implied turn-over in structure function
indicates that the characteristic time scale for optical variability of quasars 
is of the order 1 year in the rest frame. The long-term ($\ga$1 year) quasar 
variability decreases with luminosity and rest-frame wavelength similarly 
to the short-term ($\la$1 year) behavior. 

A particularly valuable result of comparing POSS and SDSS catalogs is the 
selection of candidate RR Lyrae stars, which are excellent probes of the Milky Way's 
halo structure. We demonstrated that the known halo substructures are recovered 
by the selected candidates, and discovered several new features. This method 
will eventually yield several thousand RR Lyrae candidates. 

Our study also revealed some limitations of the POSS catalogs. In particular, 
we had to limit our search for variable sources to only isolated point sources 
detected by both POSS and SDSS. Attempts to find sources detected by only one 
survey, or variable sources that are blended with another nearby source, were 
unsuccessful due to overwhelmingly large numbers of false positives. 

Despite these shortcomings, the assembled catalogs of candidate variable
sources offer a good starting point for further analysis and follow-up 
observations. For example, light curves and spectra for selected candidates
could be obtained even with telescopes of modest size. Such additional data
would help improve candidates' classification beyond 
information provided by SDSS colors. Another potentially interesting research 
direction is positional cross-correlation with catalogs obtained at other wavelengths 
(e.g. ROSAT, 2MASS, IRAS). For example, long-period variables such as Mira and 
other AGB stars are typically strong infrared emitters, and thus could be
efficiently separated from the rest of candidate variables. 

This study once again demonstrates the importance of maintaining a careful
archive of astronomical observations; the data may be valuable long after
the acquisition technology becomes obsolete.

\vskip 0.4in \leftline{Acknowledgments}

We thank Princeton University for generous financial support of this research.
 
Funding for the creation and distribution of the SDSS Archive has been provided 
by the Alfred P. Sloan Foundation, the Participating Institutions, the National 
Aeronautics and Space Administration, the National Science Foundation, the U.S. 
Department of Energy, the Japanese Monbukagakusho, and the Max Planck Society. 
The SDSS Web site is http://www.sdss.org/.

The SDSS is managed by the Astrophysical Research Consortium (ARC) for the 
Participating Institutions. The Participating Institutions are The University 
of Chicago, Fermilab, the Institute for Advanced Study, the Japan Participation 
Group, The Johns Hopkins University, Los Alamos National Laboratory, the 
Max-Planck-Institute for Astronomy (MPIA), the Max-Planck-Institute for 
Astrophysics (MPA), New Mexico State University, University of Pittsburgh, 
Princeton University, the United States Naval Observatory, and the 
University of Washington.

\newpage

\appendix{\bf Appendix A: An Overview of POSS Catalogs Used in This Work }

\vskip 0.5in
\leftline{\bf USNO catalogs}

USNO-A2.0 is a catalog of 526,280,881 stars, based on a re-reduction of the 
Precision Measuring Machine (PMM) scans of Palomar Observatory Sky Survey I 
(POSS-I; Minkowski \& Abel 1963) O and E plates, the UK Science Research Council 
SRC-J survey plates, and the European Southern Observatory ESO-R survey plates. 
For field centers with $\delta > -30^\circ$, data come from POSS-I plates, while 
data for field centers with $\delta < -35^\circ$ come from SRC-J and ESO-R plates. 
USNO-A2.0 catalog uses the ICRF as realized by the USNO ACT catalog (Urban et al. 
1997), and in addition to source coordinates, it lists the blue ($O$) and red ($E$) 
magnitudes for each object. The USNO-B catalog (\cite{Monet03}), currently 
released in version 1.0, is the next in a sequence of catalogs produced\footnote{Available from 
http://www.nofs.navy.mil} by the USNO team. It is an all-sky catalog with positions, 
proper motions, magnitudes in different optical bands (\O, \E, \J, \F, \N), and 
star/non-star estimator for aprox. 1 billion objects. Beside the first epoch surveys 
(POSS-I, ESO-R, SRC-J), it also utilizes the second epoch surveys: POSS-II
for field centers with $\delta > -30^\circ$, and SES (South Equatorial Survey) for 
$\delta < -35^\circ$  (\cite{Reid91}). It is fairly complete to $V=21$, with the claimed 
astrometric accuracy of $0.2\arcsec$  (J2000), photometric accuracy of 0.3 mag, and $85\%$ 
accuracy for distinguishing stars from non-stellar objects.

\vskip 0.5in
\leftline{\bf The Guide Star Catalog}

The GSC II (McLean et al. 2000) is an all-sky catalog based on scans of the photographic 
plates obtained by the Palomar and UK Schmidt telescopes. Schmidt plates for both the 
Northern (POSS-II) and Southern (SES) hemisphere surveys were digitized on the GAMMA 
scanning machines. Positions, magnitudes, and classifications are produced for all objects
on each plate and the data is stored in the COMPASS database. The GSC2.2 catalog is an 
all-sky, magnitude-selected export of calibrated source parameters from the COMPASS 
database, complete to $F=18.5$ mag and $J=19.5$ mag. It uses $\sim$1000 objects per 
plate for astrometric calibration, resulting in astrometric errors of $0.3\arcsec$, and 
$\sim$100 objects per plate for photometric calibration, resulting in errors in the range 
0.2--0.25 mag. The number of unique objects exported is approximately 456 million\footnote{For 
more details and access to catalogs see http://www-gsss.stsci.edu/gsc/GSChome.htm}.

\vskip 0.5in
\leftline{\bf Digital Palomar Observatory Sky Survey (DPOSS)}

The Digital Palomar Observatory Sky Survey (DPOSS; Djorgovski et al. 1998) is a 
digital version of the Second Palomar Observatory Sky Survey (POSS-II), based on the 
plate scans done at STScI, CCD calibrations done at Palomar, and processing done at 
Caltech. DPOSS consists of the original image database ($\sim$3 TB of pixel data) and 
the derived catalogs and metadata, primarily the Palomar-Norris Sky Catalog (PNSC). 
DPOSS is a survey of the northern sky ($\delta>-3^\circ$), in 3 bands (photographic 
$J,F,N$, calibrated to $g,r,i$; note that we use upper case letters for DPOSS magnitudes
to distinguish them from SDSS magnitudes), with typical limiting magnitudes $G\sim21-21.5$, 
$R\sim21$, and $I\sim19.5$ mag. Accurate star-galaxy classification is available for 
all objects to $\sim1-1.5$ mag above the detection limit\footnote{The survey and selected 
data products are publicly available from http://dposs.caltech.edu}. The initial data 
release covers the high Galactic latitudes. The final catalog is expected to contain about 
50 million galaxies and a billion stars.

\newpage

\begin{deluxetable}{lrrr}
\tablenum{1} \tablecolumns{4} \tablewidth{200pt}
\tablecaption{Best-fit Coefficients for POSS-SDSS Photometric Transformations}
\tablehead {Band & $b$ & $c$ & $\sigma$}
\startdata
 O$_a$ &  0.354   & -0.32  & 0.26 \\
 E$_a$ & -0.101   & -0.30  & 0.25 \\
 O     &  0.444   &  0.05  & 0.31 \\
 E     & -0.162   & -0.29  & 0.27 \\
 J     &  0.075   &  0.10  & 0.32 \\
 F     & -0.133   & -0.14  & 0.20 \\
 N     & -0.530   & -0.37  & 0.25 \\
 J$_g$ &  0.105   &  0.20  & 0.14 \\
 F$_g$ & -0.101   & -0.18  & 0.10 \\
 G     & -0.392   & -0.28  & 0.20 \\
 R     & -0.127   &  0.10  & 0.17 \\
\enddata
\tablenotetext{a}{For the definitions of $b$ and $c$ see eq. (1). The fourth
column ($\sigma$) lists the residual rms scatter.} 
\end{deluxetable}

\begin{deluxetable}{lcccccccccc}
\tablenum{2} \tablecolumns{11} \tablewidth{480pt}
\tablecaption{Summary of Improvements in POSS Photometry}
\tablehead {Catalog & $\sigma_O^{old}$ & $\sigma_O^{new}$ & $\sigma_E^{old}$ & $\sigma_E^{new}$ & $\sigma_J^{old}$ & $\sigma_J^{new}$ & $\sigma_F^{old}$ & $\sigma_F^{new}$ & $\sigma_N^{old}$ & $\sigma_N^{new}$}
\startdata
USNO-A2.0  & 0.260 & 0.154 & 0.248 & 0.171 &  ---  &  ---  &  ---  &  ---  &  ---  & ---   \\ 
USNO-B1.0  & 0.294 & 0.137 & 0.294 & 0.149 & 0.311 & 0.112 & 0.201 & 0.116 & 0.261 & 0.175 \\ 
GSC2.2     &  ---  &  ---  &  ---  &  ---  & 0.138 & 0.076 & 0.101 & 0.091 &  ---  & ---   \\ 
DPOSS$^b$      &  ---  &  ---  &  ---  &  ---  & 0.189 & 0.086 & 0.162 & 0.112 &  ---  & --- \\ 
\enddata
\tablenotetext{a}{``Old'' refers to rms scatter before recalibration, and ``new'' to rms
scatter after recalibration.} 
\tablenotetext{b}{$J$ and $F$ bands listed for DPOSS correspond to $G$ and $R$ bands}
\end{deluxetable}

\begin{deluxetable}{rrcrr}
\tablenum{3} \tablecolumns{6} \tablewidth{300pt}
\tablecaption{The Number of Selected Candidate Variables}
\tablehead { band  &  $|\Delta m|_{min}^a$  &  $|\Delta m/\sigma|_{min}^b$ & $m_{faint}^c$ & $N_{sel}^d$ }
\startdata
  O   &    0.35     &  2.5 &   19.0  &    25,112 \\ 
  E   &    0.35     &  2.5 &   18.0  &    16,080 \\ 
  J   &    0.20     &  2.5 &   18.0  &     6,026 \\  
  F   &    0.20     &  2.5 &   17.5  &    14,755 \\  
  G   &    0.20     &  2.5 &   19.5  &    10,824 \\  
  R   &    0.20     &  2.5 &   18.5  &     7,889 \\  
G$^e$ &    0.30     &  3.5 &   19.0  &     1,112 \\  
R$^e$ &    0.30     &  3.5 &   18.5  &     1,066 \\  
\enddata
\tablenotetext{a}{The minimum value of magnitude change}
\tablenotetext{b}{The minimum value of magnitude change normalized by the estimated photometric error}
\tablenotetext{c}{The adopted faint limit (using synthesized SDSS magnitudes)}
\tablenotetext{d}{The number of selected candidate variables in SDSS DR1 area (2099 deg$^2$)}
\tablenotetext{e}{Restricted SDSS-DPOSS selection (see Section~\ref{SDSSmulti}), magnitude limit 
corresponds to SDSS $g$ band. There are $\sim1.4\times10^6$ isolated point sources brighter than 
the same magnitude limit in the analyzed area.}
\end{deluxetable}

\begin{deluxetable}{rlrrrr}
\tablenum{4} \tablecolumns{6} \tablewidth{360pt}
\tablecaption{The distribution of variable sources in the $g-r$ vs $u-g$ diagram}
\tablehead {Region$^a$ & Name$^b$ & \% all$^c$ & \% var$_{G}^d$ & \% var$_{R}^e$ & all/var$_{G}$ }
\startdata
I   & white dwarfs &  0.13 &   0.45 &  0.09 &   3.34  \\
II  & low-z QSOs   &  0.43 &  18.6  &  3.38 &   42.8  \\
III & binary stars &  0.10 &   6.21 &  1.13 &   61.6  \\
IV  & RR Lyrae     &  0.76 &  11.3  &  9.66 &   14.9  \\
V   & blue stars   &  75.7 &  47.8  &  53.4 &   0.63  \\
VI  & red stars    &  22.9 &  16.0  &  25.6 &   0.70  \\
VII & high-z QSOs  &  0.02 &   2.43 &  7.32 &  132.7  \\
\enddata
\tablenotetext{a}{The regions boundaries are shown in Fig.~\ref{CCDs}} 
\tablenotetext{b}{An approximate description of the dominant source type} 
\tablenotetext{c}{Fraction of all SDSS sources in the region (based on run 752)} 
\tablenotetext{d}{Fraction of G-selected candidate SDSS-DPOSS variable sources in the region, restricted selection from Table 3} 
\tablenotetext{e}{Fraction of R-selected candidate SDSS-DPOSS variable sources in the region, restricted selection from Table 3} 
\end{deluxetable}

\begin{deluxetable}{rrrr}
\tablenum{5} \tablecolumns{6} \tablewidth{300pt}
\tablecaption{The Long-Term Quasar Variability}
\tablehead { band  &    rms$^a$ &  error$^b$ & $\sigma^c$ }
\startdata
               J   &    0.24     &  0.08 &    0.23   \\  
               F   &    0.23     &  0.11 &    0.20   \\  
               G   &    0.25     &  0.10 &    0.23   \\  
               R   &    0.28     &  0.14 &    0.24   \\  
               O   &    0.47     &  0.20 &    0.42   \\ 
               E   &    0.37     &  0.18 &    0.32   \\ 
\enddata
\tablenotetext{a}{The rms scatter for quasars}
\tablenotetext{b}{The rms scatter for stars with similar magnitude 
distribution as the selected quasars}
\tablenotetext{c}{An estimate for the intrinsic rms variability of quasars, $\sigma=\sqrt{rms^2-error^2}$}
\end{deluxetable}



\newpage
\clearpage


\clearpage

\clearpage

\end{document}